\documentclass[sigconf,screen,balance=false]{acmart}

\renewcommand\footnotetextcopyrightpermission[1]{}
\setcopyright{none}
\acmConference[Preprint]{}{March 5}{2026}

\usepackage{microtype}
\usepackage{bbold}
\usepackage{graphicx}
\usepackage{multirow}
\usepackage{subcaption}
\usepackage{enumitem}
\usepackage{soul}
\usepackage{color}
\usepackage{xcolor}
\usepackage{epsfig}
\usepackage{natbib}
\usepackage{acronym}
\usepackage{pifont}
\usepackage{tikz}
\usepackage[utf8]{inputenc}
\usepackage{amsmath}
\usepackage{amsfonts}
\usepackage{amsthm}
\usepackage[linesnumbered,ruled,vlined]{algorithm2e}
\usepackage[switch]{lineno}
\usepackage{xspace}
\usepackage{booktabs}
\usepackage{comment}

\usepackage{textcomp}

\usepackage{hyperref}
\hypersetup{
    colorlinks=true,  %
    linkcolor=blue,   %
    citecolor=red,    %
    filecolor=cyan,   %
    urlcolor=magenta  %
}

\definecolor{ForestGreen}{RGB}{34,139,34}

\usepackage[labelfont=bf]{caption}
\usepackage[T1]{fontenc}
\usepackage{fancyvrb}

\newcommand{\eg}{{\it e.g.}}
\newcommand{\etal}{\textit{et al}.\xspace}

\newcommand{\ie}{{\it i.e.}}
\newcommand{\vs}{\textit{vs.}\xspace}

\newcommand*\circled[1]{\tikz[baseline=(char.base)]{\node[shape=circle,draw,inner sep=1.5pt] (char) {#1};}}

\SetKwSty{text}
\SetArgSty{texttt}
\SetDataSty{}
\SetKwInput{Input}{\textcolor{blue}{\textbf{Input}}}
\SetKwInput{Output}{\textcolor{blue}{\textbf{Output}}}
\SetKwIF{If}{ElseIf}{Else}{\textcolor{blue}{\textbf{if}}}{\textcolor{blue}{\textbf{then}}}{\textcolor{blue}{\textbf{else if}}}{\textcolor{blue}{\textbf{else}}}{\textcolor{blue}{\textbf{endif}}}
\SetKwFor{For}{\textcolor{blue}{\textbf{for}}}{\textcolor{blue}{\textbf{do}}}
{\textcolor{blue}{\textbf{endfor}}}
\SetKwFunction{Length}{length}
\SetKwFor{While}{\textcolor{blue}{\textbf{while}}}{\textcolor{blue}{\textbf{do}}}
{\textcolor{blue}{\textbf{endwhile}}}
\SetKwFunction{Max}{max}
\SetKwProg{Fn}{\textcolor{blue}{\textbf{function}}}{}{}
\SetKw{Return}{\textcolor{blue}{\textbf{return}}}
\SetKwComment{Comment}{/* }{ */}

\acrodef{ML}[ML]{Machine Learning}
\newcommand{\ML}{\ac{ML}\xspace}

\acrodef{NSF}[NSF]{National Science Foundation}

\acrodef{AI}[AI]{Artificial Intelligence}
\newcommand{\AI}{\ac{AI}\xspace}

\acrodef{FL}[FL]{Federated Learning}

\acrodef{CL}[CL]{Critical Learning}

\acrodef{AC}[AC]{Attacking-Critical}

\acrodef{CAGR}[CAGR]{compound annual growth rate}

\acrodef{CCT}[CCT]{Center for Computation and Technology}

\acrodef{SLO}[SLO]{service level objective}

\acrodefplural{SLOs}[SLO]{service level objectives}

\acrodef{RL}[RL]{reinforcement learning}

\acrodef{DRL}[DRL]{deep reinforcement learning}
\newcommand{\DRL}{\ac{DRL}\xspace}

\acrodef{VM}[VM]{virtual machine}

\acrodefplural{VM}[VMs]{virtual machines}

\acrodef{ITC}[ITC]{Innovation \& Technology Commercialization}

\acrodef{DAG}[DAG]{directed acyclic graph}

\acrodefplural{DAG}[DAGs]{directed acyclic graphs}

\acrodef{SFA}[SFA]{single point authentication}

\acrodef{HPC}[HPC]{high-performance computing}

\acrodef{SBIR}[SBIR]{Small Business Innovation Research}

\acrodef{IoT}[IoT]{Internet of Things}

\acrodef{DML}[DML]{distributed machine learning}

\acrodef{GNN}[GNN]{graph neural network}

\acrodefplural{GNN}[GNNs]{graph neural networks}

\acrodefplural{NN}[NNs]{neural networks}
\newcommand{\NNs}{\acp{NN}\xspace}

\acrodef{BSR}[BSR]{backdoor success rate}

\acrodef{BTA}[BTA]{backdoor task accuracy}

\acrodef{ATT}[ATT]{App Tracking Transparency}

\acrodef{DNN}[DNN]{deep neural network}

\acrodef{KL}[KL]{Kullback–Leibler}

\acrodef{IaaS}[IaaS]{Infrastructure-as-a-Service}

\acrodef{CaaS}[CaaS]{Container-as-a-Service}

\acrodef{TRPO}[TRPO]{Trust Region Policy Optimization}

\acrodef{CPO}[CPO]{Constrained Policy Optimization}

\acrodef{PPO}[PPO]{Proximal Policy Optimization}

\acrodef{TV}[TV]{Total Variation}

\acrodef{PAC}[PAC]{Probably Approximately
Correct}

\acrodef{ACI}[ACI]{Azure Container Instances}

\acrodef{NLP}[NLP]{Natural Language Processing}
\newcommand{\NLP}{\ac{NLP}\xspace}

\acrodef{GAE}[GAE]{Generalized Advantage Estimation}

\acrodef{IS}[IS]{Importance Sampling}

\acrodef{CV}[CV]{coefficient of variance}
\newcommand{\CV}{\ac{CV}\xspace}

\acrodefplural{CV}[CVs]{coefficient of variances}

\acrodef{LLM}[LLM]{Large Language Model}
\newcommand{\LLM}{\ac{LLM}\xspace}

\acrodefplural{LLM}[LLMs]{Large Language Models}
\newcommand{\LLMs}{\acp{LLM}\xspace}

\acrodef{GNS}[GNS]{gradient noise scale}

\acrodef{SOTA}[SOTA]{state-of-the-art}
\newcommand{\SOTA}{\ac{SOTA}\xspace}

\acrodef{SOTP}[SOTP]{state-of-the-practice}

\acrodef{SSP}[SSP]{Stale Synchronous Parallel}

\acrodef{CDF}[CDF]{cumulative distribution function}
\newcommand{\CDF}{\ac{CDF}\xspace}

\acrodef{PDF}[PDF]{probability density function}

\acrodef{RPC}[RPC]{remote procedure call}

\acrodef{MoE}[MoE]{Mixture-of-Experts}
\newcommand{\MoE}{\ac{MoE}\xspace}

\acrodef{KD}[KD]{knowledge distillation}

\acrodef{DL}[DL]{Deep Learning}

\acrodef{KV}[KV]{key-value}

\acrodef{TTFT}[TTFT]{Time-To-First-Token}

\acrodef{TPS}[TPS]{Tokens-Per-Second}

\acrodef{TPOT}[TPOT]{Time-Per-Output-Token}

\acrodef{FFN}[FFN]{feed-forward network}
\newcommand{\FFN}{\ac{FFN}\xspace}
 
\acrodefplural{FFN}[FFNs]{feed-forward networks}

\acrodef{LRU}[LRU]{least recently used}

\acrodef{LFU}[LFU]{least frequently used}

\acrodef{FLOP}[FLOP]{floating point operation}

\acrodefplural{FLOP}[FLOPs]{floating point operations}
\newcommand{\FLOPs}{\acp{FLOP}\xspace}

\acrodef{ILP}[ILP]{integer linear programming}
\newcommand{\ILP}{\ac{ILP}\xspace}

\acrodef{EPM}[EPM]{expert probability map}

\acrodefplural{EPM}[EPMs]{expert probability maps}

\acrodefplural{FEP}[FEP]{fine-grained expert probability}

\acrodef{DP}[DP]{data parallelism}
\newcommand{\DP}{\ac{DP}\xspace}

\acrodef{MP}[MP]{model parallelism}
\newcommand{\MP}{\ac{MP}\xspace}

\acrodef{EP}[EP]{expert parallelism}
\newcommand{\EP}{\ac{EP}\xspace}

\acrodef{NN}[NN]{neural network}

\acrodef{EPLB}[EPLB]{Expert Parallelism Load Balancer}
\newcommand{\EPLB}{\ac{EPLB}\xspace}

\newcommand{\sys}{\textit{MoEless}\xspace}

\newcommand{\mixtral}{Mixtral-8$\times$7B\xspace}

\newcommand{\phimoe}{Phi-3.5-MoE\xspace}
\newcommand{\llama}{Llama-4-Scout\xspace}
\newcommand{\lmsys}{LMSYS-Chat-1M\xspace}
\newcommand{\sharegpt}{ShareGPT\xspace}
\newcommand{\megatron}{Megatron-LM\xspace}

\begin{document}

\date{}

\title{\sys: Efficient MoE LLM Serving with Serverless Experts}

\author{Hanfei Yu}
\authornote{Both authors contributed equally to this work.}
\email{hyu42@stevens.edu}
\affiliation{%
  \institution{Stevens Institute of Technology}
  \city{}
  \state{}
  \country{}
}

\author{Bei Ouyang}
\authornotemark[1]
\authornote{This work was conducted while Bei Ouyang was a remote research intern advised by Dr. Hao Wang at the IntelliSys Lab, Stevens Institute of Technology.}
\email{bei.ouyang@utexas.edu}
\affiliation{%
  \institution{The University of Texas at Austin}
  \city{}
  \state{}
  \country{}
}

\author{Shwai He}
\email{shwaihe@umd.edu}
\affiliation{%
  \institution{University of Maryland College Park}
  \city{}
  \state{}
  \country{}
}

\author{Ang Li}
\email{angliece@umd.edu}
\affiliation{%
  \institution{University of Maryland College Park}
  \city{}
  \state{}
  \country{}
}

\author{Hao Wang}
\email{hwang9@stevens.edu}
\affiliation{%
  \institution{Stevens Institute of Technology}
  \city{}
  \state{}
  \country{}
}

\begin{abstract}

Large Language Models (LLMs) increasingly adopt Mixture-of-Experts (MoE) architectures to scale efficiently under stringent resource constraints.
However, MoE’s sparse activation causes severe expert load imbalance, where a few experts become stragglers while others remain underutilized, leading to inflated inference latency and cost. 
Existing solutions assume static, serverful model deployments, limiting expert elasticity and often incurring costly expert swapping or degraded output quality. 

We present \sys, an efficient serverless MoE serving framework that mitigates expert load imbalance via elastic expert execution. 
\sys leverages lightweight, layer-aware predictors to estimate incoming expert load distributions and proactively identify stragglers. 
We design optimized scaling and placement strategies to improve function locality, GPU utilization, and cross-expert load balance. 
\sys is prototyped on top of Megatron-LM and deployed on an eight-GPU testbed.
Experiments with open-source MoE models and real-world workloads show that \sys reduces inference latency by 43\% and inference cost by 84\% compared to state-of-the-art solutions.

\end{abstract}

\maketitle %

\section{Introduction}
\label{sec:intro}

\LLMs have revolutionized \NLP research and driven breakthroughs across diverse application domains, such as content generation~\cite{dai2024neural,achiam2023gpt,brown2020language,radford2019language}, information retrieval and recommendation~\cite{lin2024data,zhao2024let}, and \AI-assisted decision-making~\cite{nam2024using,li2024go,jiang2024lilac}.
To mitigate the extreme training costs, modern \LLMs increasingly adopt the \MoE architecture~\cite{jiang2024mixtral,snowflake-arctic,yang2024qwen2,xai-grok,dai2024deepseekmoe,abdin2024phi} as their core design.
\MoE layers replace the conventional \FFN layers in Transformer blocks~\cite{vaswani2017attention} with a gating network and a collection of experts, where only a small subset is activated during computation.
This design significantly reduces the number of \FLOPs, enabling \MoE-based \LLMs to deliver comparable or even superior performance to dense \LLMs at a fraction of the training cost~\cite{dai2024deepseekmoe,yang2024qwen2,jiang2024mixtral}.

\begin{figure}[t]
    \centering
   \includegraphics[width=\linewidth]{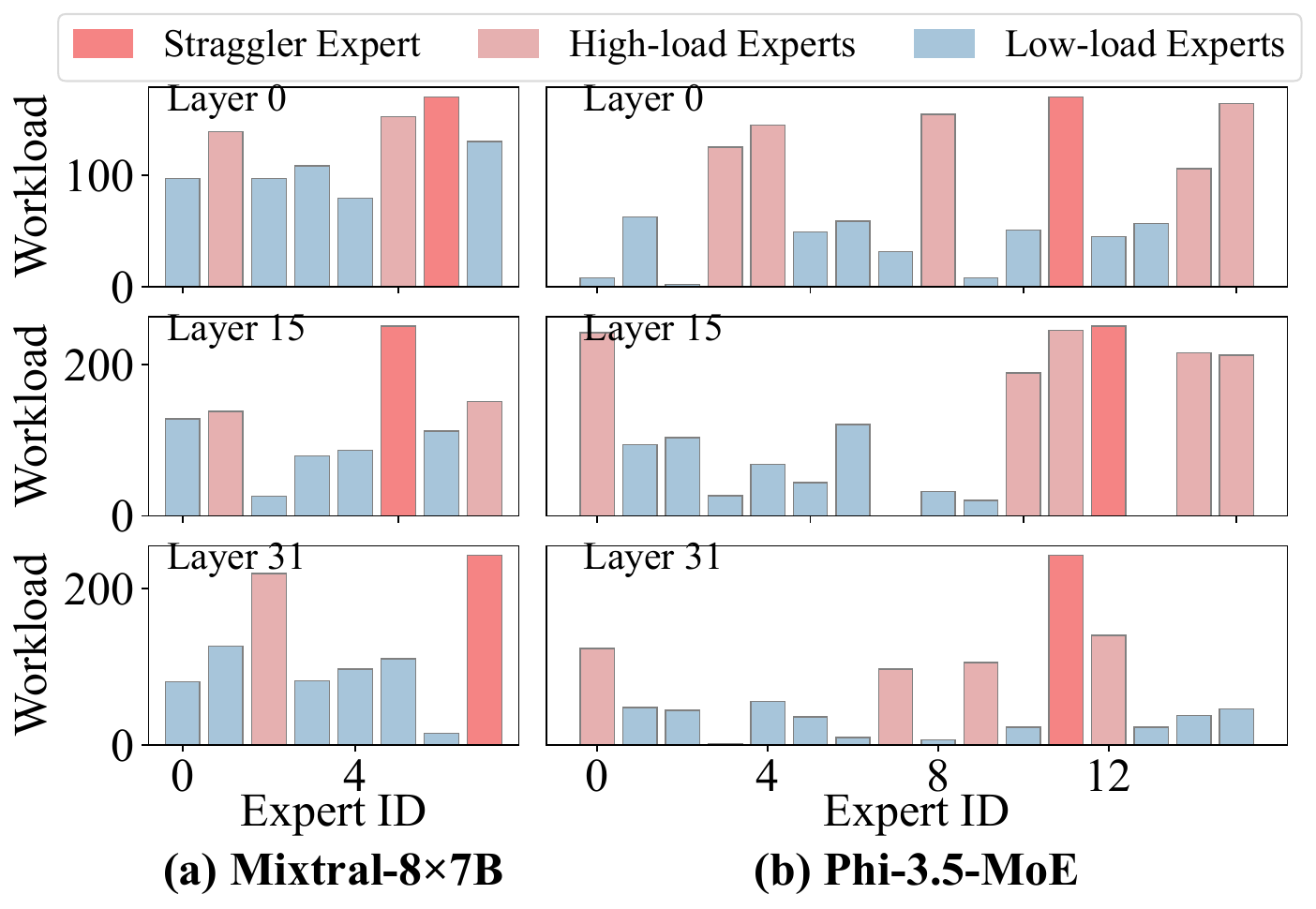}
   \caption{Expert load imbalance across layers for different MoE models and datasets: (a) \mixtral on \sharegpt and (b) \phimoe on \lmsys.}
    \label{fig:workload_imbalance}
\end{figure}

Due to the immense model scale, serving \MoE-based \LLMs requires distributed deployment under the \EP paradigm~\cite{liu2024deepseek}.
However, \textit{expert load imbalance} has been identified as a fundamental challenge in distributed serving~\cite{liu2024deepseek,he2022fastermoe,liu2025netmoe}, where certain experts become highly popular and receive overwhelming loads compared to others.
Figure~\ref{fig:workload_imbalance} illustrates the expert load imbalance in two representative \MoE models, \mixtral and \phimoe, across two real-world datasets.
Such imbalance leads to the \textit{expert straggler} problem~\cite{he2026capacity}, which severely increases inference latency and serving cost in \MoE serving.

Expert load balancing~\cite{liu2024deepseek,li2023accelerating,he2026capacity} has been extensively explored to mitigate the inherent load imbalance among experts in \MoE serving.
However, existing approaches assume static resource configurations on serverful infrastructures, resulting in either costly real-time expert swapping~\cite{liu2024deepseek,he2022fastermoe,li2023accelerating} with limited effectiveness, or lossy expert re-routing~\cite{he2026capacity} that compromises generation quality.
Achieving expert load balance without generation performance loss requires fine-grained, elastic, and accurate expert scaling in \MoE serving.

\textit{Serverless computing} offers a promising alternative and has emerged as a new computing paradigm for modern \AI infrastructures.
Serverless \ML inference and \LLM serving have gained significant attention from both academia~\cite{fu2024serverlessllm,sui2024pre,sui2025serverlesslora,zeng2025medusa,lou2026hydraserve} and industry~\cite{aws-sagemaker,azure-ml,aws-bedrock,nv-nim}.
In this paper, we propose \sys, an efficient serverless \MoE serving framework that mitigates expert load imbalance and accelerates \MoE inference via serverless experts.
\sys decouples experts from \MoE models under the \EP paradigm and integrates them with serverless functions to enable scalable and elastic execution.
To proactively identify expert stragglers, we design lightweight predictors that accurately estimate upcoming expert load distributions with layer awareness.
Based on the predicted load characteristics, \sys dynamically scales expert replicas to eliminate stragglers and balance workloads across both the expert and GPU levels, thereby minimizing inference latency.
Furthermore, we design optimized expert placement strategies to maximize function locality and GPU utilization while reducing all-to-all communication overheads in \EP deployments.
In summary, we make the following contributions:
\begin{itemize}
    \item We propose \sys, an efficient serverless \MoE serving framework that accelerates inference by mitigating expert load imbalance via serverless experts.
    \item We design layer-aware and lightweight expert load predictors that accurately estimate incoming expert load distributions across different layers.
    \item We develop dynamic expert scaling and placement strategies that efficiently balance workloads across experts and GPUs to eliminate straggler problems.
    \item We prototype \sys on top of \megatron~\cite{shoeybi2019megatron}, deploy it on an eight-GPU testbed, and conduct extensive evaluation against \SOTA approaches.
\end{itemize}

\section{Background and Motivation}

\subsection{Large Language Model Serving}

\begin{figure}[t]
    \centering
    \includegraphics[width=\linewidth]{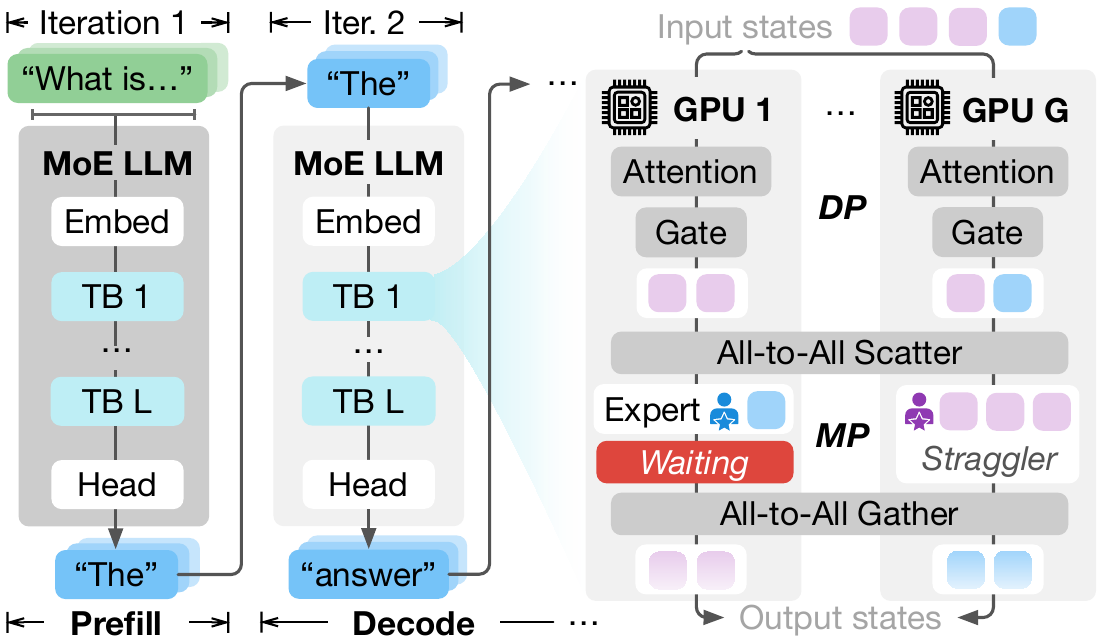}
    \caption{Illustration of serving Mixture-of-Experts (MoE) based Large Language Models under expert parallelism, where tokens are routed by per-layer gate networks to a sparse set of experts distributed across GPUs.
    Expert load imbalance triggers inefficient resource provisioning (\eg, over-scaling hot experts or under-utilizing cold ones), thereby increasing serving cost.
    \textbf{Embed}: embedding layer, \textbf{TB}: Transformer Block, \textbf{Head}: language modeling head, \textbf{Attention}: attention layer, \textbf{Gate}: gate networks, \textbf{DP}: data parallelism, \textbf{MP}: model parallelism.}
    \label{fig:bg-ep}
\end{figure}

\LLMs have been extensively studied and deployed in both industrial production~\cite{patel2024splitwise,agrawal2024taming,kwon2023efficient,qiu2025modserve} and academic research~\cite{lee2024infinigen,liu2024cachegen,du2025prefillonly,zhong2024distserve}.
Modern \LLMs typically adopt the Transformer decoder-only architecture~\cite{vaswani2017attention}, which operates in two consecutive stages: \textit{prefill} and \textit{decode}.
Figure~\ref{fig:bg-ep} illustrates the auto-regressive inference process, where an \LLM processes an input prompt batch and incrementally generates new tokens.
During the prefill stage, the model processes all input prompt sequences in parallel and outputs the first new tokens for the batch in a single \textit{iteration}.\footnote{An iteration refers to one inference step that generates a new token. The iteration time represents the end-to-end step latency.}
This stage is typically compute-intensive because it runs full attention over the entire prompt length, and it also initializes the key-value (KV) cache that will be reused for subsequent generation.
In contrast, the decode stage is latency-sensitive and often memory-bandwidth-bound: at each iteration, the model generates one new token per sequence while attending to all previously generated tokens via the KV cache.
As the context grows, the amount of KV-cache reads increases monotonically, making per-iteration performance increasingly dominated by cache access and communication rather than raw compute.
To improve throughput, serving systems commonly batch requests and dynamically adjust batch composition across iterations; however, batching also introduces trade-offs between utilization and tail latency, especially when sequences have diverse prompt lengths and output lengths.
In the decode stage, the model generates one new token per sequence in each iteration until all responses are completed, and the overall end-to-end latency is therefore determined by both the iteration time and the total number of decode iterations required.

\subsection{Mixture of Experts and Expert Parallelism}

The \MoE architecture has emerged as a dominant paradigm for scaling \LLMs beyond dense models, enabling hyper-scale parameters without proportionally increasing computation~\cite{shazeer2017outrageously,du2022glam,fedus2022switch,dai2024deepseekmoe,jiang2024mixtral}.
As illustrated in Figure~\ref{fig:bg-ep}, \MoE-based \LLMs replace the standard \FFN layer in each Transformer block~\cite{vaswani2017attention} with an \MoE layer, which consists of a gating network and multiple expert networks.
Within each block, the attention module first computes token-level attention~\cite{vaswani2017attention} from the input hidden states.
The gating network then routes tokens to specific experts, and each token activates only its assigned expert for computation.
Compared with traditional dense \LLMs, \MoE-based models activate only a subset of parameters during training and inference, substantially reducing computation while achieving superior generation quality with a comparable total parameter count.

Given the extreme scale, \MoE-based \LLMs are typically served through distributed deployment to meet their intensive computational and memory demands.
Existing \MoE serving systems~\cite{li2023accelerating,shi2024schemoe,liu2024deepseek} employ \EP, which integrates both \DP and \MP as shown in Figure~\ref{fig:bg-ep}.
In each Transformer block, non-expert modules (\eg, attention layers and gating networks) are replicated across GPUs for parallel data processing, while experts are uniquely distributed across GPUs to accommodate their large memory footprints (\eg, each expert in \mixtral occupies 0.33~GB of GPU memory).
Consequently, two all-to-all communication operations are required between non-expert modules and experts to realize token-to-expert assignments: a \textit{scatter} operation that distributes tokens to their designated experts, and a \textit{gather} operation that collects and reorders the outputs from experts~\cite{liu2024deepseek,he2022fastermoe,li2023accelerating}.

\subsection{Expert Load Imbalance in MoE Serving}
\label{subsec:bg-expert-load-imbalance}

\begin{figure}[t]
    \centering
    \includegraphics[width=0.95\linewidth]{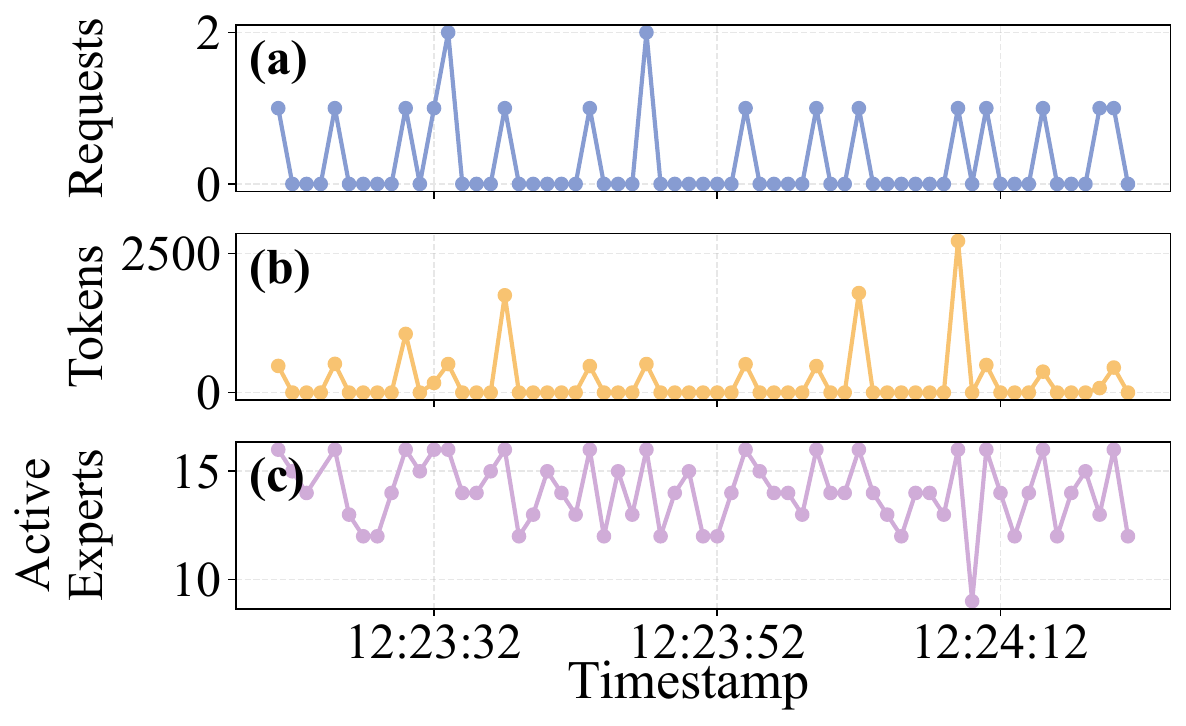}
    \caption{Serving \phimoe on \lmsys using Azure LLM traces: (a) request arrivals, (b) aggregated token loads, and (c) total number of active experts.}
    \label{fig:real_trace}
\end{figure}

Expert load imbalance poses a fundamental challenge for \MoE serving under \EP~\cite{he2026capacity,ma2025moe,li2023accelerating}.
Figure~\ref{fig:workload_imbalance} illustrates the workload distribution across three representative layers from two \MoE models, \mixtral~\cite{jiang2024mixtral} and \phimoe~\cite{abdin2024phi}, evaluated on two real-world datasets, \sharegpt~\cite{sharegpt} and \lmsys~\cite{zheng2024lmsys}.
Prior studies~\cite{hwang2023tutel,he2026capacity,balmau2025accelerating} have consistently observed that expert popularity is highly skewed, with certain experts receiving disproportionately higher loads than others.
GPUs hosting these overloaded experts take significantly longer to complete their computations, leading to the \textit{straggler problem}~\cite{he2026capacity}, where lightly loaded experts must wait for overloaded ones to finish.
Moreover, such imbalance exacerbates all-to-all communication latency in \EP deployments, as GPUs with popular experts handle larger data transfers, further amplifying the straggler effect.
Consequently, expert load imbalance can substantially increase \MoE inference latency, elevating inference cost and degrading overall serving performance.

In \MoE serving, the dynamic expert demands arise from a combination of workload-level and model-level factors:

\textbf{Varying request arrivals and token loads.} 
Figure~\ref{fig:real_trace}(a) shows the request arrivals of the real-world Azure \LLM inference traces~\cite{patel2024splitwise,stojkovic2025dynamollm}.
We replay the peak traffic observed around noon from the traces.
During \MoE serving, varying request arrivals naturally result in dynamic usage of experts.
Figure~\ref{fig:real_trace}(b) shows the total request token loads aggregated over the same traces, where we batch and sum the tokens of requests within the same second.
With varying input tokens, \MoE models unavoidably assign different loads to their experts.

\textbf{Intrinsically skewed expert popularity in \MoE.}
Existing research~\cite{he2026capacity,zhou2022mixture,balmau2025accelerating} has extensively demonstrated the highly skewed expert popularity, which stems from the \MoE architecture and the differences between prefill and decode stages.
To illustrate the dynamic expert demands, we replay the same Azure \LLM traces to serve \phimoe with \megatron~\cite{shoeybi2019megatron} using \lmsys.
Experimental details can be found in \S\ref{subsec:eval-setup}.
Figure~\ref{fig:real_trace}(c) shows the fluctuation in the number of active experts over time.

\begin{figure}[t]
    \centering
    \includegraphics[width=0.95\linewidth]{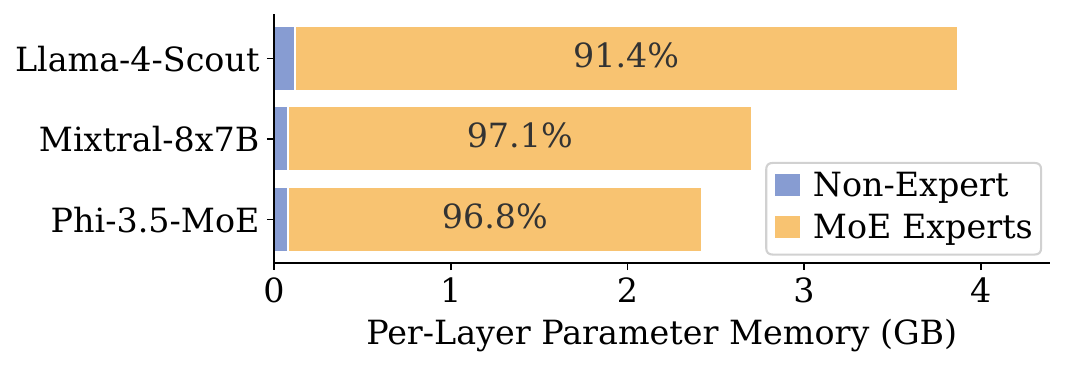}
    \caption{Per-layer parameter memory breakdown of three \MoE models. Expert parameters dominate the per-layer memory footprint, accounting for 91.4\%--97.1\% of total layer memory.}
    \label{fig:bg-per-layer-mem}
\end{figure}

\begin{figure}[t]
    \centering
    \begin{subfigure}[t]{0.49\linewidth}
        \centering
        \includegraphics[width=\linewidth]{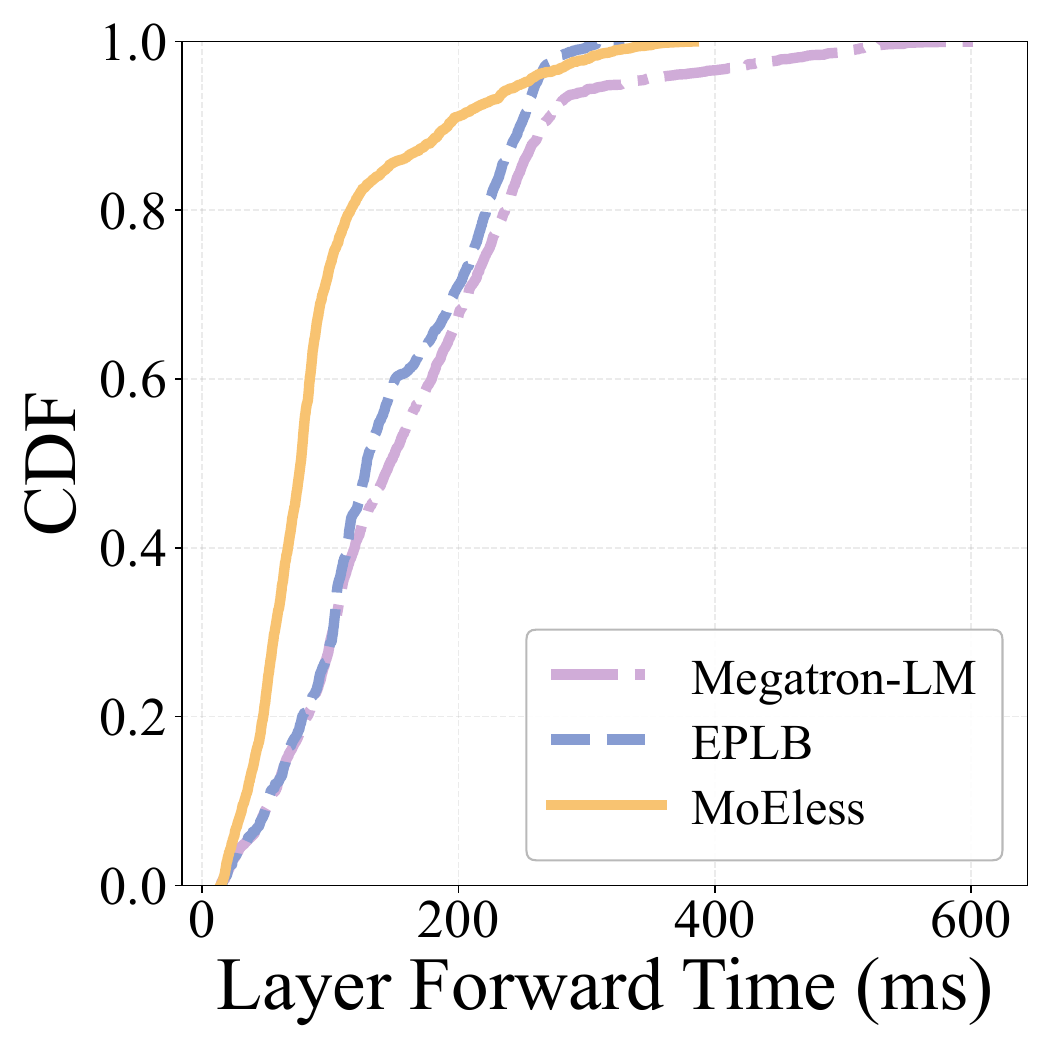}
        \caption{MoE layer forward time.}
        \label{fig:bg-serverless-latency}
    \end{subfigure}
    \hfill
    \begin{subfigure}[t]{0.49\linewidth}
        \centering
        \includegraphics[width=\linewidth]{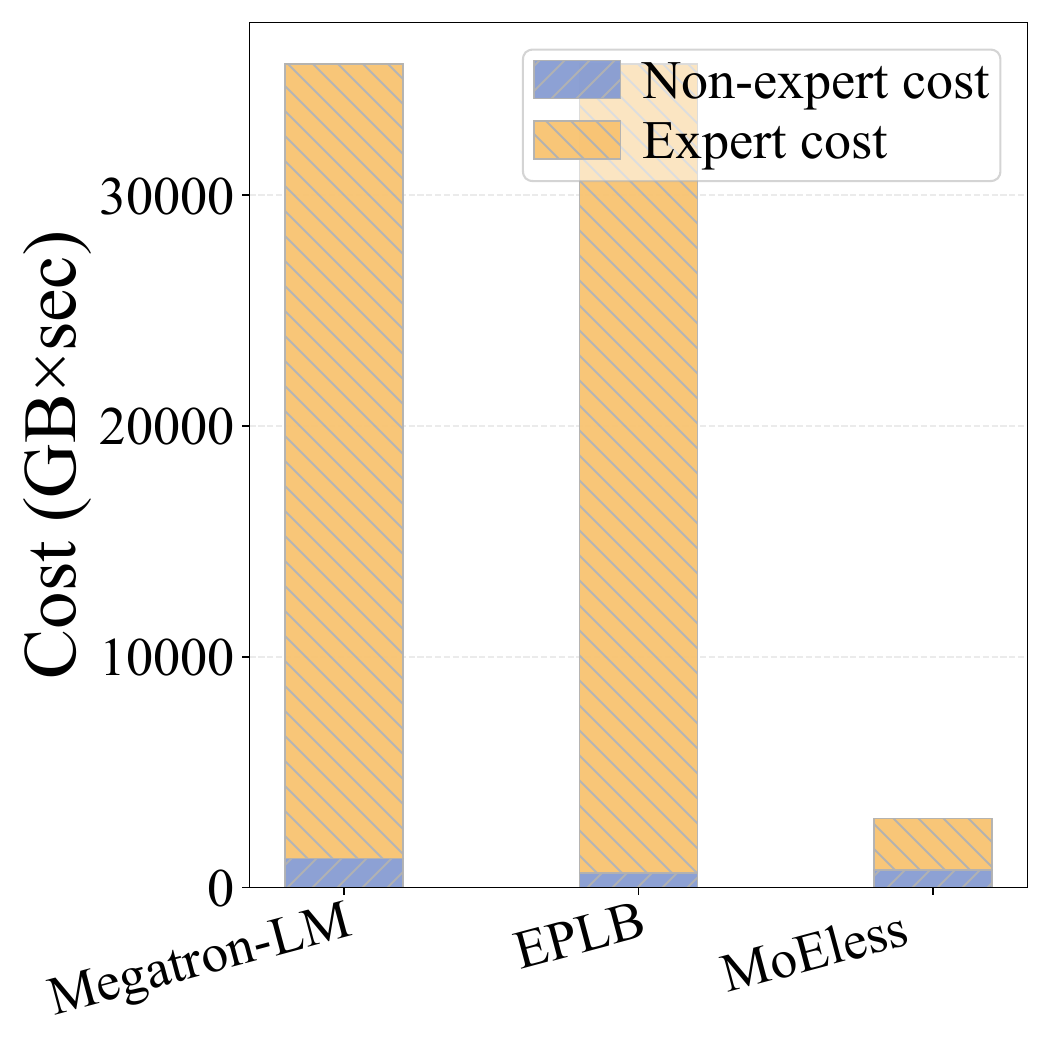}
        \caption{Total inference cost.}
        \label{fig:bg-serverless-cost}
    \end{subfigure}
    \caption{Inference performance of three approaches when serving \phimoe on \sharegpt.}
    \label{fig:bg-serverless}
\end{figure}

\subsection{Motivating Serverless MoE Serving}
\label{subsec:bg-serverless-moe}

To mitigate expert load imbalance in \MoE serving, existing solutions~\cite{liu2024deepseek,he2022fastermoe,wang2023prophet,li2023accelerating} selectively replicate popular experts for distributing the loads.
However, they rely on a fixed configuration of expert scaling on serverful infrastructures.
For example, \EPLB~\cite{liu2024deepseek} from DeepSeek assumes a fixed number of expert replicas on a fixed number of devices.
Figure~\ref{fig:bg-per-layer-mem} illustrates why expert replication is expensive by showing the per-layer parameter memory breakdown of \mixtral, \phimoe, and \llama.
Across three representative \MoE models, expert parameters account for 91.4\%--97.1\% of the per-layer memory footprint.
Consequently, replicating popular experts substantially increases GPU memory consumption, limiting the number of replicas that can be provisioned on serverful infrastructures.
As a result, serverful \MoE serving fails to satisfy dynamically changing expert demands (\S\ref{subsec:bg-expert-load-imbalance}).

\textit{Serverless computing} has emerged as a transformative paradigm in modern \AI infrastructures, offering agile scalability, pay-as-you-go pricing, and simplified management to enable scalable \ML inference~\cite{ali2020batch,zhang2019mark,sui2024pre,guo2022hydrozoa} and elastic \LLM serving~\cite{fu2024serverlessllm,zeng2025medusa,lou2026hydraserve}.
In contrast, traditional serverful serving systems (\eg, \megatron~\cite{shoeybi2019megatron}) deploy experts on fixed compute nodes with static \EP configurations, often suffering from severe straggler effects and expert load imbalance.
Existing serverful expert load balancing approaches~\cite{liu2024deepseek,he2022fastermoe} attempt to mitigate this issue by swapping low-usage experts with replicas of popular ones.
However, their expert scalability and elasticity remain constrained by fixed resource allocations, resulting in inflated per-layer latency and higher inference costs.
In contrast, a serverless \MoE architecture dynamically scales experts on demand, effectively eliminating hidden stragglers and achieving balanced workloads across experts and GPUs.
As shown in Figure~\ref{fig:bg-serverless}, the serverless design significantly reduces both \MoE layer forward latency and inference cost over serverful baselines.
\section{Overview}

\subsection{Objectives and Challenges}

We design \sys to achieve three goals:

\textbf{Accelerate \MoE inference by mitigating expert load imbalance.}
Existing \MoE-based \LLMs inevitably experience expert load imbalance from both workload and model perspectives, causing the expert straggler problem~\cite{he2026capacity} that prolongs inference latency.
We aim to alleviate the expert load imbalance and accelerate \MoE inference by eliminating expert stragglers.

\textbf{Enable expert scalability via serverless computing.}
Existing serverful \MoE serving frameworks~\cite{he2022fastermoe,liu2024deepseek,li2023accelerating} rely on swapping out low-usage experts for popular ones limited by fixed resource allocations.
By leveraging serverless computing, we aim to unlock expert scalability and elasticity to efficiently accommodate dynamic expert demands.

\textbf{Minimize \MoE inference cost with serverless experts.}
Since expert computations dominate the inference latency and cost~\cite{jiang2024mixtral,abdin2024phi,llama4,liu2024deepseek}, we aim to reduce the inference cost with serverless experts.

We must address three challenges to realize the goals:

\textbf{How to integrate serverless computing into existing \MoE serving?}
Unlike dense \LLMs, the scale of \MoE-based \LLMs is prohibitively large, making it infeasible to encapsulate the entire model within serverless functions~\cite{fu2024serverlessllm,zeng2025medusa,lou2026hydraserve,hu2025deepserve}.
Therefore, our design must be fine-grained and generic to seamlessly integrate with existing \MoE serving frameworks.

\textbf{How to accurately predict expert load distributions?}
To enable proactive expert load balancing, we must effectively capture and predict dynamic expert load distributions in advance, providing accurate guidance for asynchronous expert management and scaling decisions.

\textbf{How to efficiently scale and place expert functions?}
Given the predicted expert loads, we must construct execution plans that efficiently scale serverless expert functions and place them across devices to maximize GPU utilization, function locality, and minimize communication overheads.

\subsection{Architecture and Workflow}
\label{subsec:architecture}

\begin{figure}[t]
    \centering
    \includegraphics[width=0.9\linewidth]{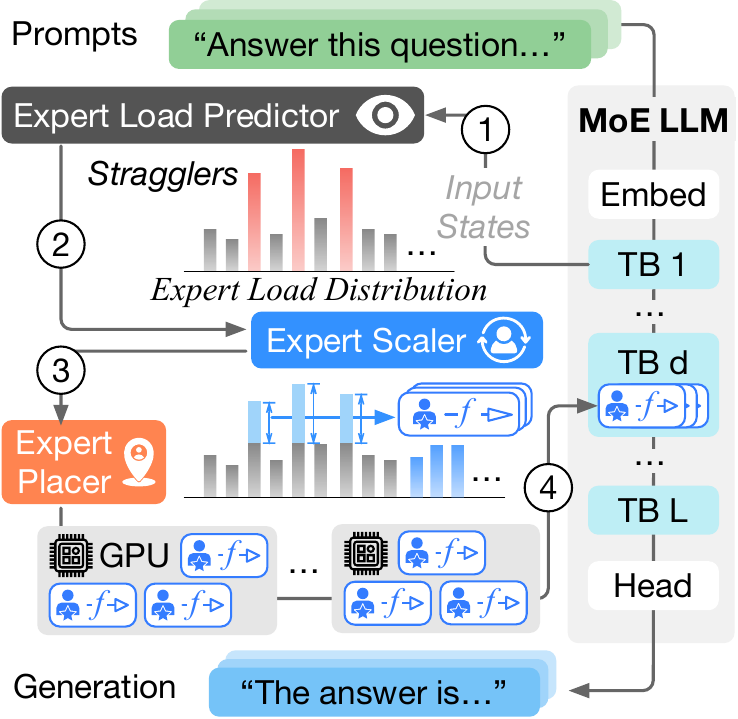}
    \caption{The architecture and workflow of \sys.}
    \label{fig:overview-arch}
\end{figure}

Unlike existing serverless \LLM serving approaches~\cite{fu2024serverlessllm,zeng2025medusa,lou2026hydraserve,hu2025deepserve}, which encapsulate the entire \LLM into serverless functions, we decouple the experts from the \MoE model and package them as independent functions, while keeping the non-expert modules in \DP.
This design offers two key advantages.
First, since expert computation dominates inference latency and cost, packaging experts as functions maximizes the benefits of serverless execution while avoiding unnecessary coordination among non-expert modules.
Second, because experts are inherently decoupled and follow all-to-all communication patterns, the stateless nature of serverless functions can be naturally hidden under \EP, ensuring seamless integration with existing \MoE serving workflows.

Figure~\ref{fig:overview-arch} shows the architecture and workflow of \sys, consisting of three main components: \textbf{Expert Load Predictor}, \textbf{Expert Scaler}, and \textbf{Expert Placer}.
\sys serves \MoE models in four steps:

\noindent \textbf{Step {\Large \circled{\small 1}}: Expert load prediction.}
Unlike existing expert selection predictors~\cite{eliseev2023fast,song2024promoe,hwang2024pre,yu2025taming,xue2024moe}, which operate at the single-request level, our \textit{Expert Load Predictor} accurately estimates expert load distributions across request batches and identifies batch-level expert stragglers for each \MoE layer (\S\ref{subsec:design-predictor}).
The prediction results are then passed to the \textit{Expert Scaler} for subsequent expert management.

\noindent \textbf{Step {\Large \circled{\small 2}}: Expert scaling.}
The \textit{Expert Scaler} receives the predicted expert loads and determines expert scaling decisions under inference budget constraints (\S\ref{subsec:design-scaler}).
It first analyzes expert stragglers and sets a target forward latency for each predicted layer.
Next, it trims excessive straggler loads to meet the latency target and allocates additional replicas to handle the overflow.
The resulting expert scaling plan is then forwarded to the \textit{Expert Placer} for deployment.

\noindent \textbf{Step {\Large \circled{\small 3}}: Expert placement.}
With \EP, each expert instance must be assigned to a GPU for execution.
The \textit{Expert Placer} generates an optimized GPU placement strategy based on prior expert states and hardware characteristics, given the new scaling plan (\S\ref{subsec:design-placer}).
The placement is designed to maximize function locality, improve GPU utilization, and minimize communication overhead (\eg, expert migration between GPU and CPU, or all-to-all data transfers).

\noindent \textbf{Step {\Large \circled{\small 4}}: Expert serving.}
The inference process consists of one iteration in the prefill stage and multiple iterations in the decode stage.
For each \MoE layer in every iteration, we evenly distribute each expert’s load across its replicas, achieving dynamic load balancing and eliminating stragglers through parallel processing.

\subsection{Problem Formulation}
\label{subsec:problem-formulation}

Following prior \MoE serving research~\cite{yu2025taming}, we consider serving a \MoE-based \LLM composed of $L$ \MoE layers on a cluster of $G$ homogeneous GPUs.
Each \MoE layer consists of one gating network and $E$ experts. 
The model processes a batch of request prompts $B$, resulting in a total expert workload (\ie, token count) of $W$.
Let $[L] := \{1, \ldots, l, \ldots, L\}$ denote the set of layers and $[E] := \{1, \ldots, e, \ldots, E\}$ the set of experts in each layer. 
Each request prompt undergoes multiple iterations across the prefill and decode stages.
During each iteration, we must make two decisions: how many replicas of each expert to instantiate and on which GPU each replica should be placed.
Let $[R^{(i,l,e)}] := \{1, \ldots, r^{(i,l,e)}, \ldots, R^{(i,l,e)}\}$ denote the set of replicas of expert $e$ at layer $l$ during iteration $i$, where $l \in [L]$, $e \in [E]$, and $i \in B$.
Let $p^{(i,l,e)}_{r,g} \in \{0,1\}$ denote whether replica $r^{(i,l,e)}$ is placed on GPU $g \in [G]$.

Since experts in an \MoE model typically share the same parameter size, we assume expert memory footprints $M_e$ to be homogeneous.
The expert workload $W_{l,e}$ is evenly divided among its replicas, yielding per-replica load $W_{l,e,r} := \frac{W_{l,e}}{R^{(i,l,e)}}$.
The processing time of a replica scales linearly with its assigned workload:
$T_{l,e,r} := \alpha \cdot W_{l,e,r}$, where $\alpha$ is a processing coefficient.
Each GPU incurs additional communication latency due to all-to-all scatter and gather operations.
Since the input and output data of an \MoE layer typically have the same size (hidden dimension), the one-time communication time of GPU $g$ is given by 
$T_g := \beta \cdot \sum_{\{p^{(i,l,e)}_{r,g}=1\}} W_{l,e,r}$, where $\beta$ is a communication coefficient.
Each layer's forward time mainly consists of the expert processing time and two rounds of all-to-all communication.

Given the definitions, we optimize two objectives: total inference latency $T$ and cost $C$.
The total inference latency is defined as
\begin{align*}
    T := \sum_{i \in B} \sum_{l \in [L]} \Big(\max_{\{e,r\}} (T_{l,e,r}) + 2 \cdot \max_{\{g\}} (T_g) + T_{\text{misc}}\Big),
\end{align*}
where $T_{\text{misc}}$ is a constant for non-\MoE latency.
The total inference cost is computed as the product of GPU memory consumption and inference latency aggregated over all iterations:
\begin{align*}
    C := &\sum_{i \in B} \sum_{l \in [L]} \big{\{} \big{[} \big{(} \max_{\{e,r\}} (T_{l,e,r}) + 2 \cdot \max_{\{g\}} (T_g) \big{)} \cdot  \\
    &\sum_{e \in [E]} \sum_{R^{(i,l,e)}} M_{e} \big{]} + T_{\text{misc}} \cdot M_{\text{misc}} \big{\}},
\end{align*}
where $M_{\text{misc}}$ is a constant for non-\MoE memory footprints.

Finally, we formulate the expert load balancing problem as a multi-objective \ILP optimization:
\begin{align*}
    &\min_{\{r^{(i,l,e)}, ~p^{(i,l,e)}_{r,g}\}} ~(T, C), \notag \\
    &\mathrm{s.t.} \sum_{\{p^{(i,l,e)}_{r,g}=1\}} M_{e} \leq M_g, \quad \forall i \in B, ~\forall g \in [G]. \label{eq:constraint-1}
\end{align*}
The objectives are minimizing the total inference latency and cost while ensuring the total memory footprint of expert replicas of any GPU $g$ satisfies the available memory $M_g$.
Solving this \ILP is NP-hard~\cite{cormen2022introduction}, and real-world workloads exhibit dynamic expert demands that further complicate the problem. 
Therefore, we opt for a heuristic-based design for \sys.

\section{Design}

\subsection{Expert Load Predictor}
\label{subsec:design-predictor}

\begin{figure}[t]
    \centering
    \includegraphics[width=\linewidth]{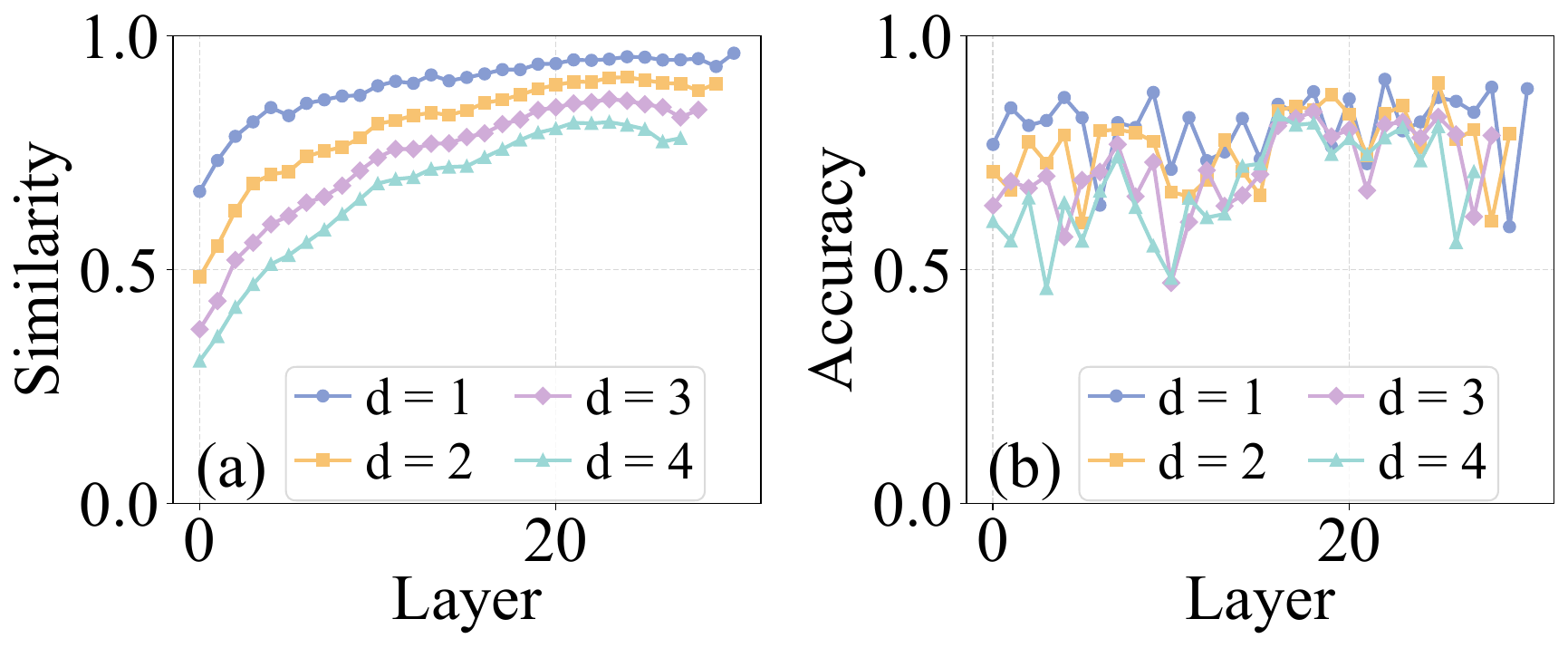}
    \caption{Characterizing \phimoe on \lmsys: (a) cosine similarity of gate network inputs, and (b) expert load prediction accuracy across layers with different prediction distances.}
    \label{fig:design-layer-aware}
\end{figure}

\begin{figure}[t]
    \centering
    \includegraphics[width=\linewidth]{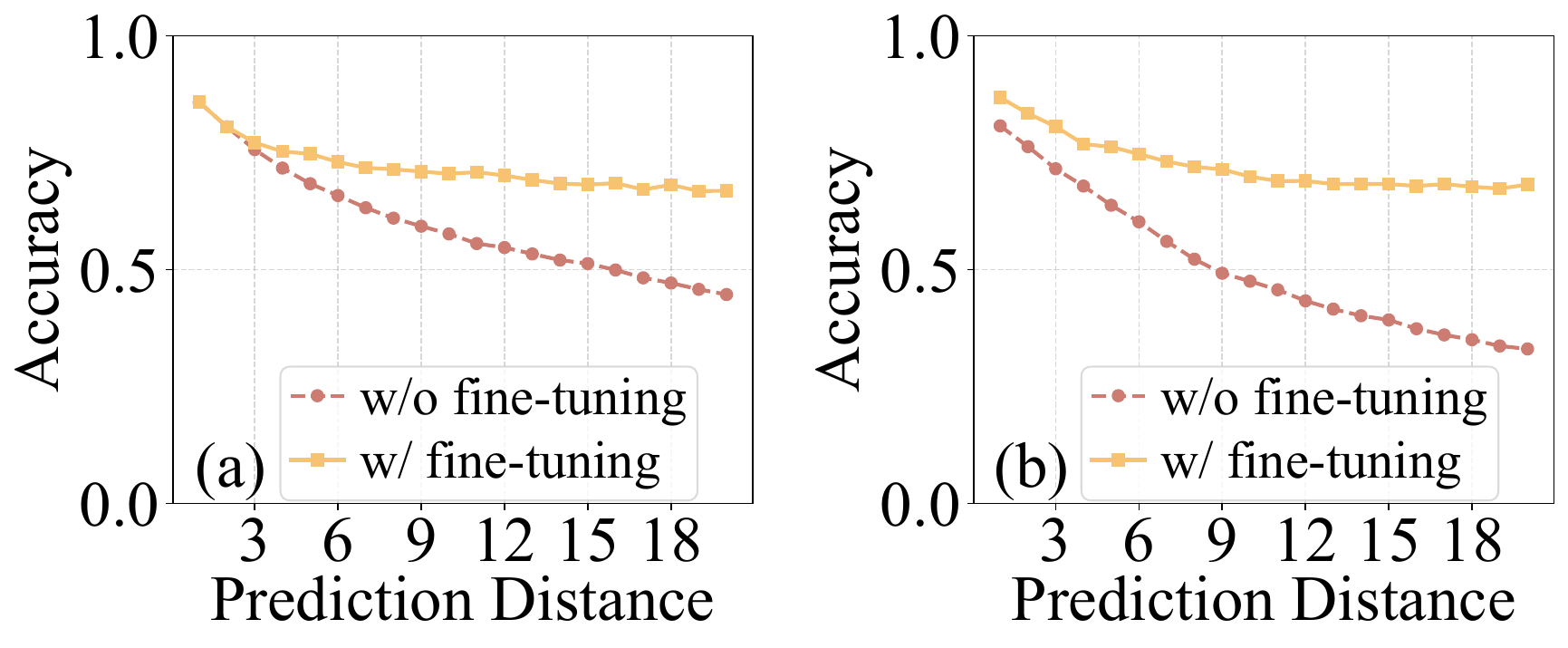}
    \caption{Expert load prediction accuracy on \lmsys with and without fine-tuning at different prediction distances: (a) \mixtral, and (b) \phimoe.}
      \label{fig:avg_accuracy_comparison}
\end{figure}

For each \MoE layer, ideal expert scaling requires waiting for the gate network outputs to reveal the actual expert loads.
However, due to the synchronization overheads of on-demand scaling and placement~\cite{he2022fastermoe,liu2025netmoe}, prior studies have explored asynchronously predicting and prefetching experts ahead of time with a prediction distance~\cite{song2024promoe,yu2025taming,xue2024moe,eliseev2023fast,zhang2025daop}.
The prediction distance $d$ refers to the number of layers in advance that scaling and placement decisions are made before the corresponding layer activates its experts.
An optimal prediction distance should fully overlap the prediction, scaling, and placement overheads with the ongoing inference process.
To this end, we design the Expert Load Predictor to accurately identify overloaded expert stragglers and mitigate them in upcoming expert executions asynchronously.

\textbf{Speculative prediction.}
Due to the common use of residual connections~\cite{he2016deep} in Transformer architectures, the hidden states output by each Transformer block remain highly similar to their inputs.
Leveraging this property, we use the input hidden states of the $l$-th layer as input to the $(l+d)$-th layer's gate network to speculate its expert load distribution $W_{l+d} := \{w^{(l+d, 1)},\allowbreak \ldots,\allowbreak w^{(l+d, e)},\allowbreak \ldots,\allowbreak w^{(l+d, E)}\}$, where $d$ is the prediction distance and $w_{l+d, e}$ is the token count for expert $e \in [E]$.
Figure~\ref{fig:design-layer-aware}(a) illustrates the cosine similarity of gate network inputs between Layers $l$ and $l+d$, where $d \in [1, 4]$, showing consistently high similarity across layers.
Figure~\ref{fig:avg_accuracy_comparison} further presents the average expert load prediction accuracy across all layers at different prediction distances for two models.
Intuitively, a larger prediction distance offers greater opportunity to overlap asynchronous expert operations, but at the cost of reduced prediction accuracy.

\textbf{Gate network fine-tuning with layer awareness.}
Existing approaches either directly reuse the original gate networks as predictors, leading to unsatisfactory prediction accuracy~\cite{eliseev2023fast}, or train large external predictors from scratch, introducing substantial computational overhead~\cite{song2024promoe}.
In contrast, we \textit{replicate} and \textit{fine-tune} the original gate networks as our predictors, preserving their inherent knowledge of expert selection while improving prediction accuracy over larger prediction distances.
Our key observation is that \textit{not all layers require the same degree of fine-tuning}.
Figure~\ref{fig:design-layer-aware}(a) shows that early layers have lower input similarity across gate networks, while later layers maintain higher input similarity and yield more reliable predictions.
Figure~\ref{fig:design-layer-aware}(b) presents the prediction accuracy across layers under different prediction distances.
Early layers tend to exhibit more unstable expert load distributions and lower accuracy, whereas later layers are more stable and predictable. 
This observation aligns with prior research~\cite{ziyin2025formation,chen2023layer}, which indicates that early layers are generally more plastic and less stable in their learning dynamics.
Leveraging this observation, we first profile each layer’s prediction accuracy prior to deployment and define a target threshold $h$ (\eg, 80\%).
Layers with accuracy below $h$ are selectively fine-tuned until they exceed the threshold.
Figure~\ref{fig:avg_accuracy_comparison} shows our layer-aware fine-tuning consistently improves prediction accuracy across varying prediction distances.

\subsection{Expert Scaler}
\label{subsec:design-scaler}

\begin{algorithm}[t]
\caption{Expert Scaling}
\label{algo:expert-scaler}

\Input{
Predicted expert loads $W_l$,
expert memory $M_e$, 
per-layer memory cap $M_{\text{cap}}$, 
CV threshold $V$
}
\Output{Replica sets $\{[R^{(l,e)}]\}$ for $\forall e \in [E]$ in $l$}

Initialize replica count $R^{(l,e)} \gets 1$, $\forall e \in [E]$\;
Initialize allocated memory $M_{\text{alloc}} \gets 0$\;

\While{$M_{\text{alloc}} < M_{\text{cap}}$ and $\texttt{CV}(W_l) > V$}{
    Select straggler $e^\ast \gets \arg\max_{e \in [E]} W_l$\,
        $W_l \gets W_l - \{w^{(l,e^{\ast})}_{r} \mid r \in [R^{(l,e^\ast)}]\}$\;
    Add one replica $R^{(l,e^\ast)} \gets R^{(l,e^\ast)} + 1$,
        $M_{\text{alloc}} \gets M_{\text{alloc}} + M_{e^\ast}$\;
    Split the load $w^{(l,e^{\ast})}_{r} \gets w^{(l,e^{\ast})}_{r} / R^{(l,e^\ast)}$,
        $W_l \gets W_l + \{w^{(l,e^{\ast})}_{r} \mid r \in [R^{(l,e^\ast)}]\}$\;
}

\Return{$\{[R^{(l,e)}]\}$}

\end{algorithm}

Upon receiving the predicted expert load distribution $W_l$ for Layer $l$, \sys's Expert Scaler determines the set of replicas $[R^{(l,e)}] := \{1, \ldots, r^{(l,e)}, \ldots, R^{(l,e)}\}$ for each expert $e \in [E]$.
As shown in Algorithm~\ref{algo:expert-scaler}, we employ a greedy heuristic to iteratively cut off the expert stragglers with high loads and converge to a balanced load distribution for each layer.
For each layer $l \in [L]$, we first initialize all experts with a single instance and a per-layer memory cap $M_{\text{cap}}$.
Then, we repeatedly identify the most overloaded expert using a max heap, add a replica to that expert, and evenly split its load until either the \CV of expert loads falls below the threshold $V$ (\eg, \CV $\leq 0.2$), or the per-layer memory cap $M_{\text{cap}}$ is reached.
This process is performed iteratively across all layers, ensuring that each layer independently achieves a balanced expert load distribution within its allocated memory budget.

\subsection{Expert Placer}
\label{subsec:design-placer}

After the number of replicas for each expert in a layer is determined, the Expert Placer decides where these replicas should be placed across available GPUs.  
We aim to minimize placement overheads while maintaining balanced GPU loads across the system.  
Our key objectives are two-fold:

\textbf{Reuse expert replicas for warm-starts.}  
If a previously placed expert replica is kept alive on a GPU, we immediately reuse it to avoid data transfer and initialization delays (\ie, function warm-starts~\cite{sui2024pre,sui2025serverlesslora,lou2026hydraserve,fu2024serverlessllm}).  
Thus, we eliminate expert cold-start overheads with pre-warming and keep-alive techniques~\cite{yu2024rainbowcake,roy2022icebreaker,shankar2020serverless,sui2024pre} from serverless computing.

\textbf{Balance per-GPU loads.}  
Since each GPU's computation and communication times are proportional to its aggregated expert load, balancing these loads minimizes straggler effects and ensures efficient parallel execution among GPUs.  
Hence, we distribute replicas across GPUs to achieve balanced utilization while satisfying per-GPU capacity constraints via a classic Join-the-Shortest-Queue algorithm~\cite{gupta2007analysis}.

As shown in Algorithm~\ref{algo:expert-placer}, we first initialize the per-GPU load tracker and placement matrix.  
Then, we select the most-loaded expert replica from all replicas, checking whether any alive replicas can be reused from the previous placement results $\{[R'^{(l,e)}]\}$.  
If reuse is not possible, we greedily assign the most-loaded replica to the GPU with the lowest current aggregated load.
After assignment, the GPU load is updated to reflect the new placement, and the process continues until all replicas are assigned.

\begin{algorithm}[t]
\caption{Expert Placement}
\label{algo:expert-placer}

\Input{
Replica set $\{[R^{(l,e)}]\}$ for $\forall e \in [E]$,
expert loads $W_l$,
GPU set $G$,
last placement results $\{[R'^{(l,e)}]\}$
}
\Output{$P_l := \{p^{(l,e)}_{r,g} \mid e \in [E], r \in [R^{(l,e)}], g \in [G]\}$}

Initialize per-GPU loads $W_g \gets \{0\}^G$, $\forall g \in [G]$\;
Initialize placement matrix $P_l \gets \{0\}^{|P_l|}$, $\forall p \in P_l$\;

\While{$W_l \neq \emptyset$}{
    Select most-loaded $r^\ast \gets \arg\max_{r \in [R^{(l,e)}]} W_l$,
        $W_l \gets W_l - w^{(l,e)}_{r^{\ast}}$\;
    \If{$r^\ast \in \{[R'^{(l,e)}]\}$ and $r^\ast \in g, \forall g \in [G]$}{
        Select warm-start $g^\ast \gets g$;
    }
    \Else{
        Select least-loaded $g^\ast \gets \arg\min_{g \in [G]} W_g$\;
    }
    Update placement $p^{(l,e)}_{r^\ast,g} \gets 1$, $W_{g} \gets W_{g} - w_{g}$\;
    Update loads $w_{g} \gets w_{g} + w^{(l,e)}_{r^\ast}$, $W_{g} \gets W_{g} + w_{g}$\;
}

\Return{$P_l$}
\end{algorithm}

\section{Implementation}

We prototype \sys on top of the \megatron framework~\cite{shoeybi2019megatron}.
The implementation details of each component are described as follows.

\textbf{Implementing predictors.}
We implement the predictors as lightweight \NNs that share the same architecture and parameter size as the \MoE model’s gate networks, using PyTorch~\cite{pytorch}.
To eliminate prediction overheads, we invoke predictors asynchronously from the main model computation using dedicated CUDA streams~\cite{cuda-runtime-api}.
For each Transformer layer, a separate CUDA stream is launched to execute the predictor on the current hidden states concurrently with the \MoE layer’s forward pass.
This design ensures that prediction is fully overlapped with computation, introducing no blocking or additional latency.

\textbf{Fine-tuning predictors.}
We collect input hidden states and the corresponding gate network outputs from each \MoE layer to construct the fine-tuning dataset.
The dataset is split into training and testing subsets with a 7:3 ratio.
For each layer, we fine-tune multiple predictors with varying inter-layer prediction distances for flexible deployment.
The fine-tuning process is lightweight, completing within five minutes on a single GPU.
To further improve training efficiency, we parallelize fine-tuning across all layers.

\textbf{Scaling and placing serverless experts.}
We support two types of expert scaling:
(1) \textit{intra-GPU scaling}, which replicates or removes expert weights within a single GPU, and
(2) \textit{inter-GPU scaling}, which adjusts the number of GPUs that host expert replicas.
Both expert scaling and placement are implemented asynchronously to minimize inference delay.
We use the NCCL~\cite{nccl} library for all-to-all communication between serverless experts and the \megatron framework.
We integrate \megatron's \EP with Docker containers~\cite{merkel2014docker}, where experts are distributed across GPU-enabled containers for execution.
Additionally, we adopt standard pre-warming and keep-alive techniques~\cite{yu2024rainbowcake,roy2022icebreaker,shankar2020serverless,sui2024pre} from serverless computing to eliminate expert cold-start latency.

\textbf{System configuration.}
Our prototype system determines its parameters (\eg, prediction distance and load-balancing thresholds) and container configurations primarily through offline profiling, rather than dynamic runtime adaptation across different models and datasets. 
We leave more advanced runtime optimization mechanisms to future work.

\begin{table}[t]
\centering
\caption{Characterizations of \MoE models used in the evaluation.}
\label{tab:moe_model_specs}
\resizebox{\columnwidth}{!}{
    \begin{tabular}{lccc}
    \toprule
    \multirow{2}{*}{\textbf{MoE Model}} &
    \textbf{Parameters} & \textbf{Experts Per Layer} & \textbf{Num. of} \\
    & \textbf{(active / total)} & \textbf{(active / total)} & \textbf{Layers} \\
    \midrule
    Mixtral-8×7B & 12.9B / 46.7B & 2 / 8  & 32 \\
    Phi-3.5-MoE  & 6.6B / 42B    & 2 / 16 & 32 \\
    Llama-4-Scout & 17B / 109B   & 1 / 16 & 48 \\
    \bottomrule
    \end{tabular}
}
\end{table}

\section{Evaluation}

\begin{figure}[t]
    \centering
    \includegraphics[width=\linewidth]{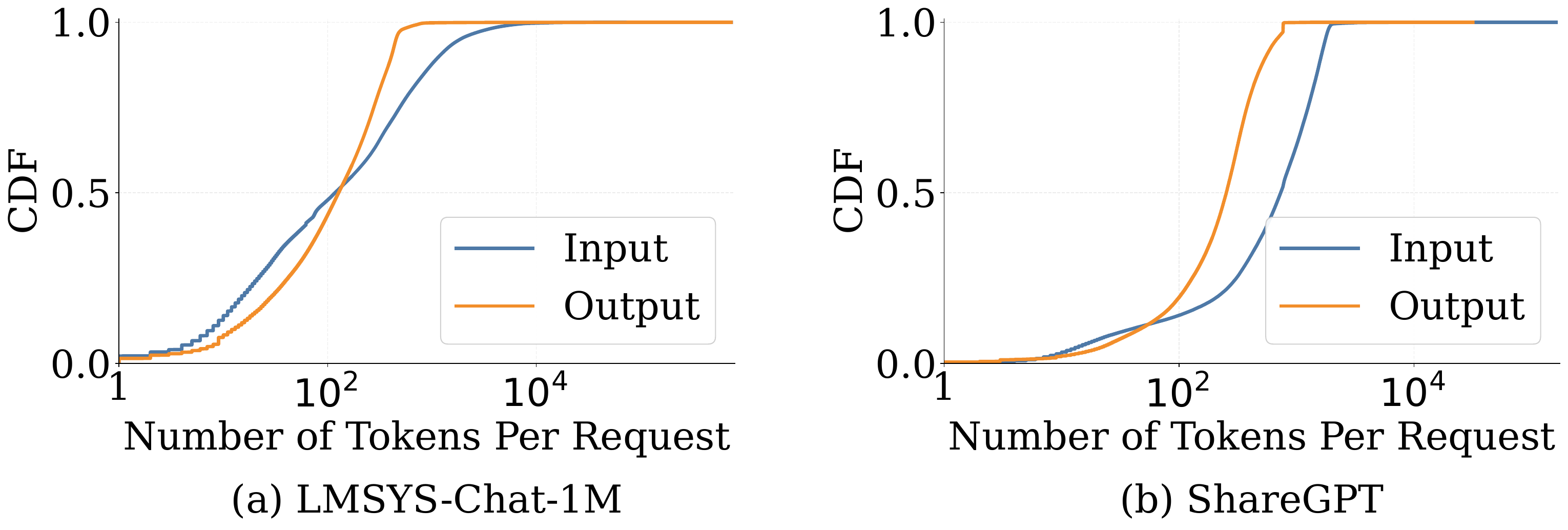}
    \caption{Characterization of \lmsys and \sharegpt datasets.}
    \label{fig:eval-datasets}
\end{figure}

\begin{figure}[t]
    \centering
    \includegraphics[width=\linewidth]{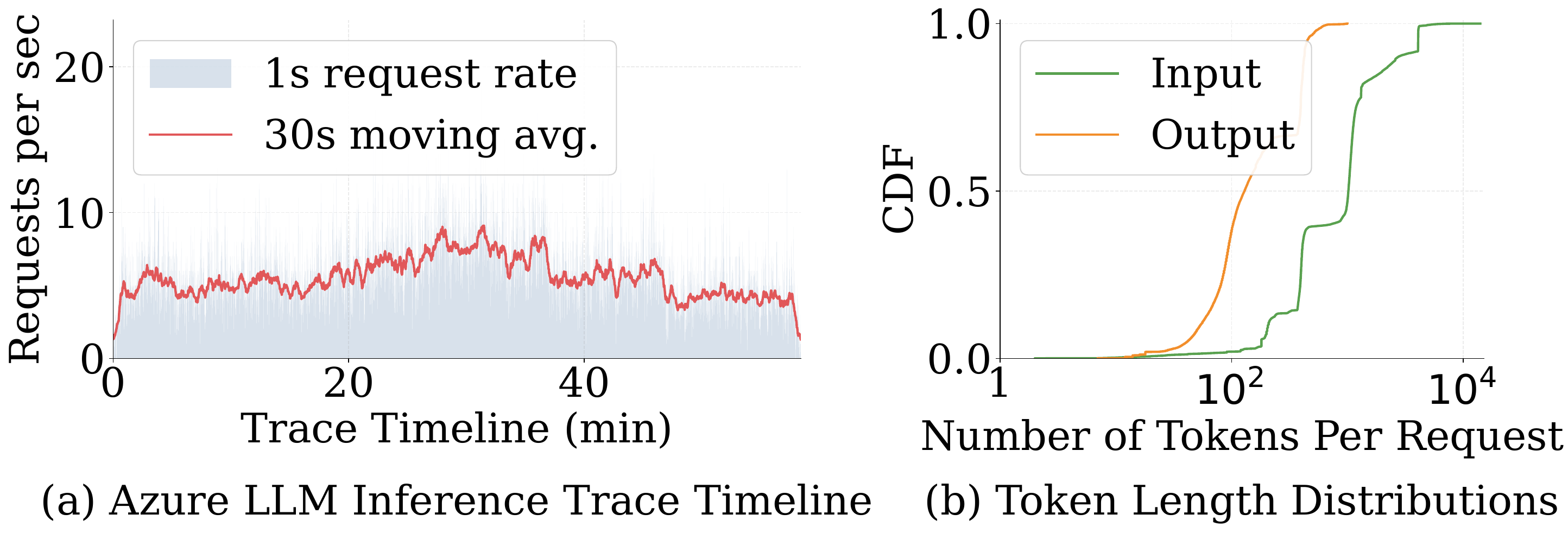}
    \caption{Characterization of Azure LLM inference traces.}
    \label{fig:eval-azure-trace}
\end{figure}

This section conducts extensive experiments to evaluate \sys, including the experimental setup (\S\ref{subsec:eval-setup}), overall performance against \SOTA baselines (\S\ref{subsec:eval-overall}), effectiveness of the expert load predictors (\S\ref{subsec:eval-prediction}), sensitivity analysis (\S\ref{subsec:eval-sensitivity}), ablation study (\S\ref{subsec:eval-ablation}), and system overheads of \sys (\S\ref{subsec:eval-overheads}).

\subsection{Experimental Setup}
\label{subsec:eval-setup}

We describe the details of experimental setup for evaluation.

\textbf{Models.}
We evaluate our system using three representative \MoE-based \LLMs: \mixtral~\cite{jiang2024mixtral}, \phimoe~\cite{abdin2024phi}, and \llama~\cite{llama4}. 
Table~\ref{tab:moe_model_specs} characterizes the three \MoE models, including the number of parameters, \MoE layers, and experts per layer, respectively.
Together, these models span recent architectural scales and expert configurations, allowing evaluation across diverse \MoE designs.

\textbf{Datasets.}
We employ two real-world prompt datasets widely used in \LLM evaluations, \lmsys \cite{zheng2024lmsys} and \sharegpt \cite{sharegpt}, as the request corpus.
Figure~\ref{fig:eval-datasets} compares their token length distributions.
While both datasets span several orders of magnitude, \lmsys generally contains longer input prompts and output responses than \sharegpt, resulting in more computationally intensive inference requests.
Evaluating on both datasets therefore captures a broad spectrum of serving workloads with diverse request lengths and generation costs.
During evaluation, requests are randomly sampled from the datasets, whereas the arrival timestamps and input/output token counts are replayed from the Azure production traces.

\begin{figure*}[t]
    \centering
    \begin{subfigure}[t]{0.328\linewidth}
        \centering
        \includegraphics[width=\linewidth]{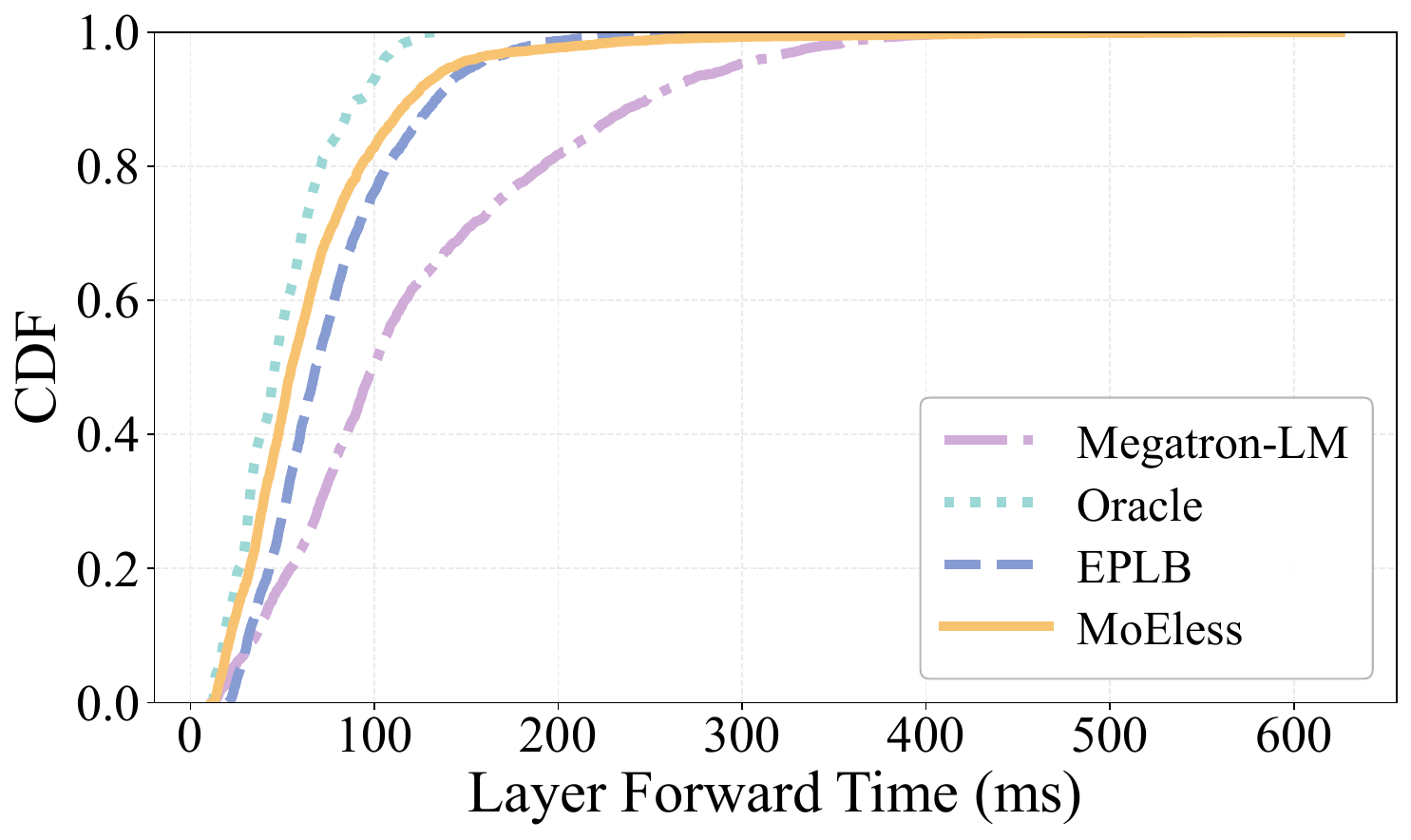}
        \caption{\mixtral.}
    \end{subfigure}
    \hfill
    \begin{subfigure}[t]{0.328\linewidth}
        \centering
        \includegraphics[width=\linewidth]{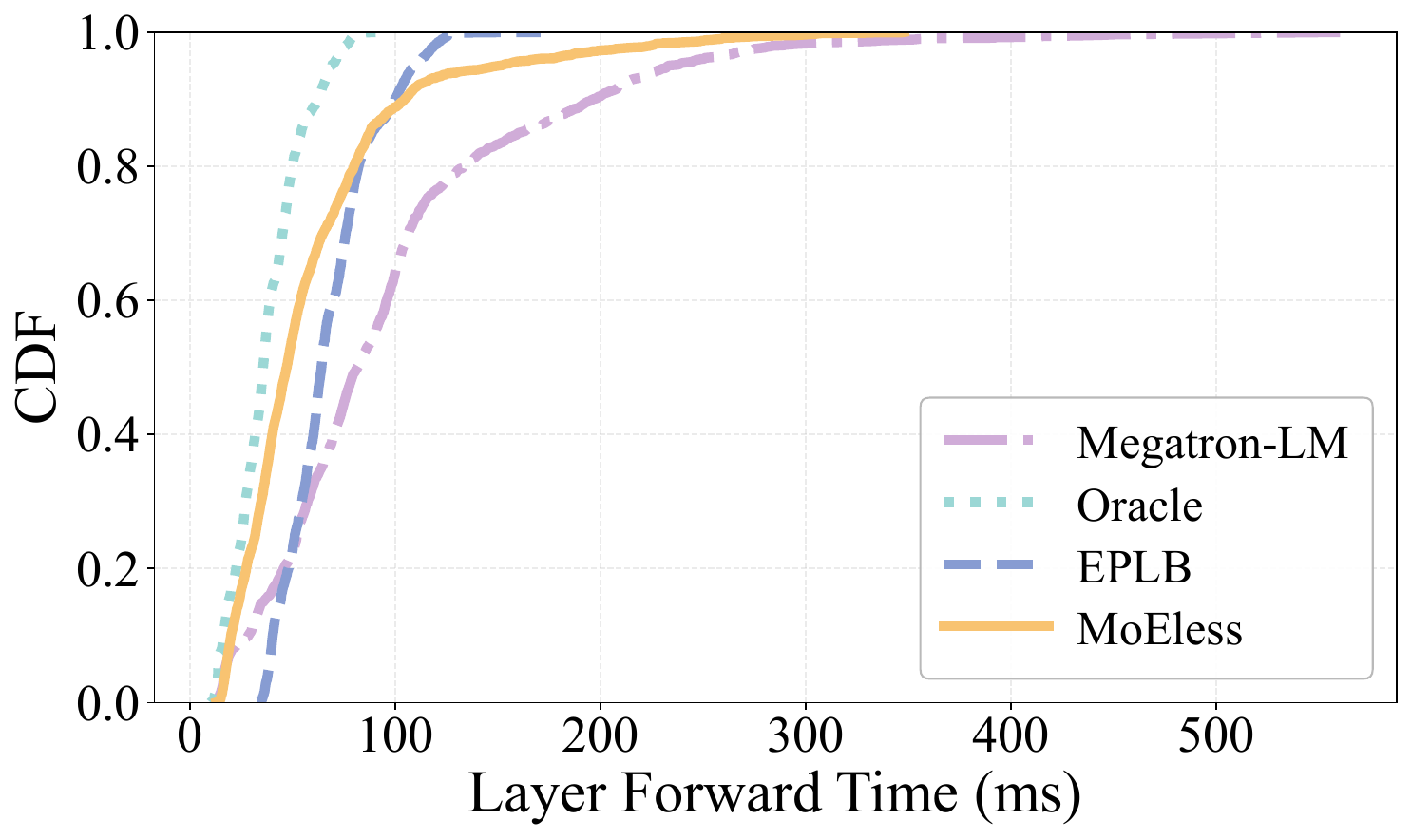}
        \caption{\phimoe.}
    \end{subfigure}
    \hfill
    \begin{subfigure}[t]{0.328\linewidth}
        \centering
        \includegraphics[width=\linewidth]{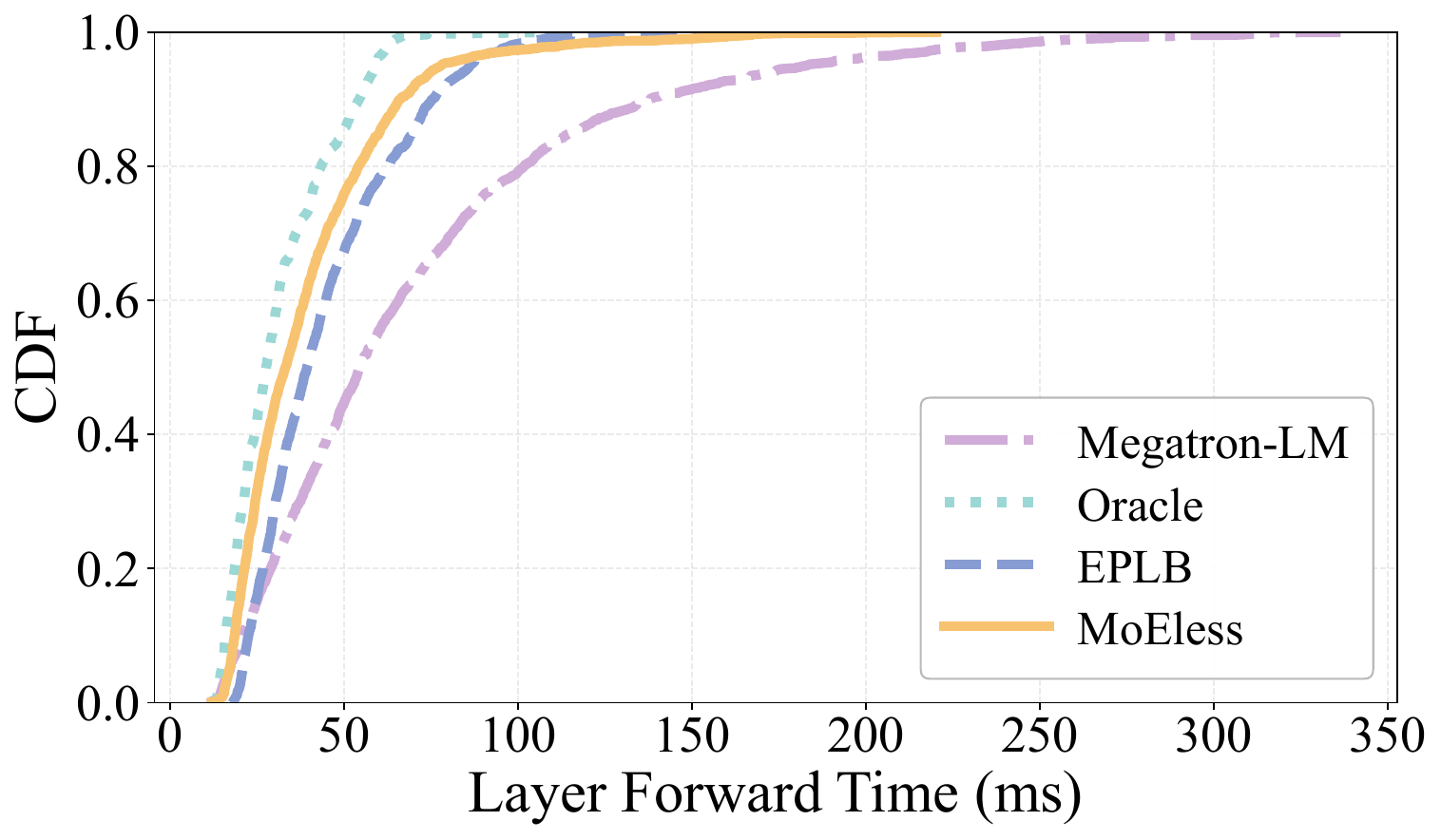}
        \caption{\llama.}
    \end{subfigure}
    \caption{MoE layer forward time of four approaches across three models on \lmsys.}
    \label{fig:eval-layer-lmsys}
    \vspace{6pt}
\end{figure*}

\begin{figure*}[t]
    \centering
    \begin{subfigure}[t]{0.328\linewidth}
        \centering
        \includegraphics[width=\linewidth]{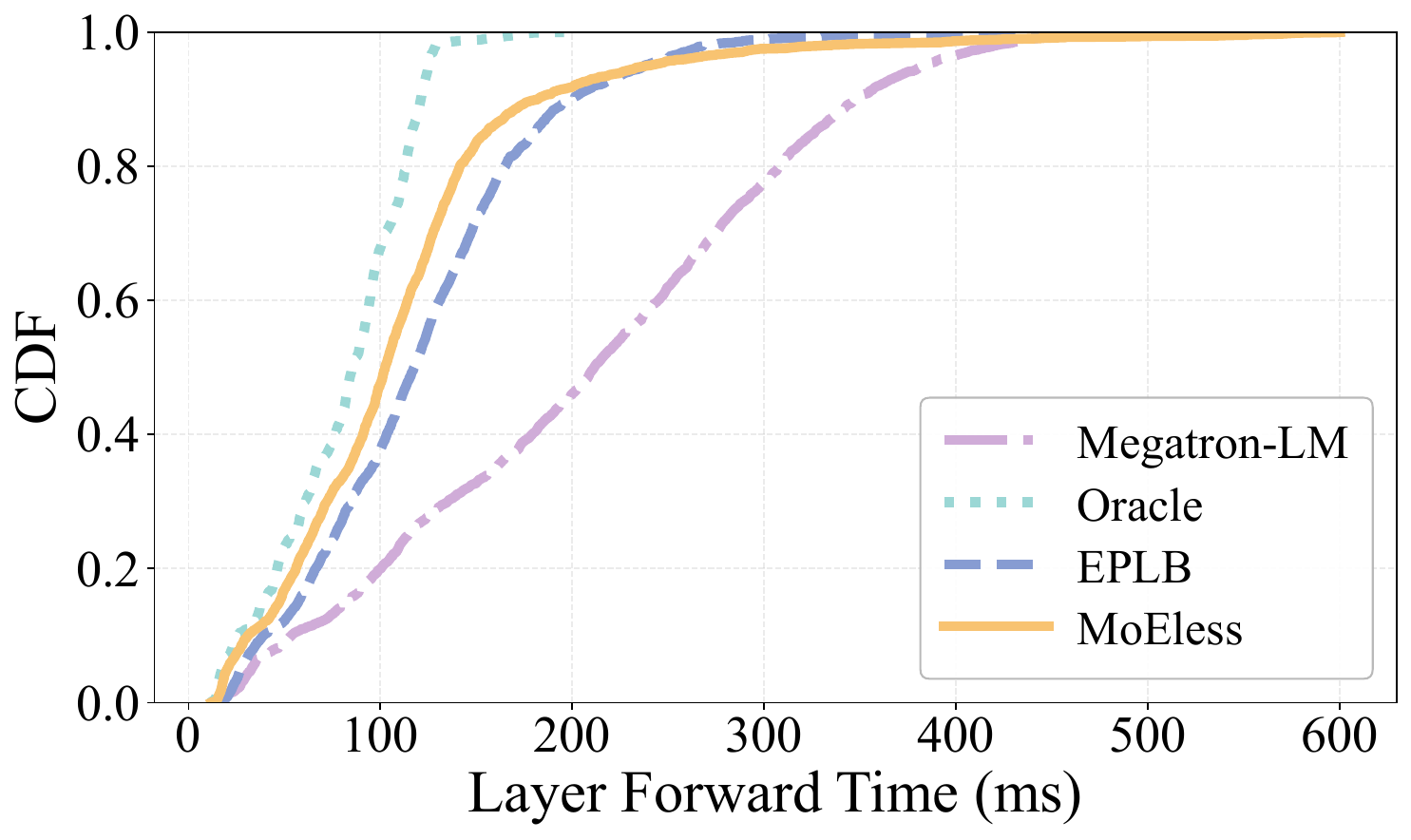}
        \caption{\mixtral.}
    \end{subfigure}
    \hfill
    \begin{subfigure}[t]{0.328\linewidth}
        \centering
        \includegraphics[width=\linewidth]{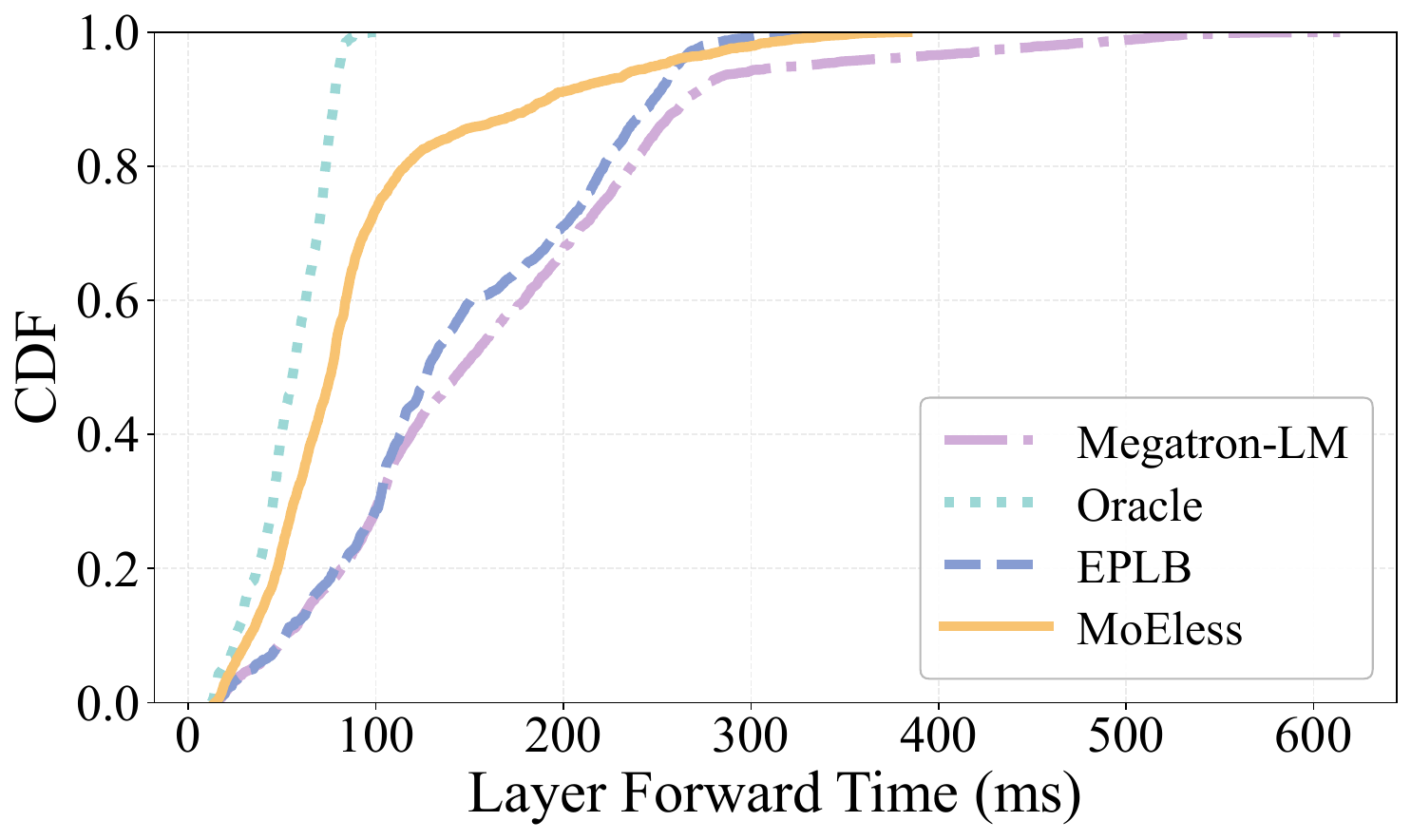}
        \caption{\phimoe.}
    \end{subfigure}
    \hfill
    \begin{subfigure}[t]{0.328\linewidth}
        \centering
        \includegraphics[width=\linewidth]{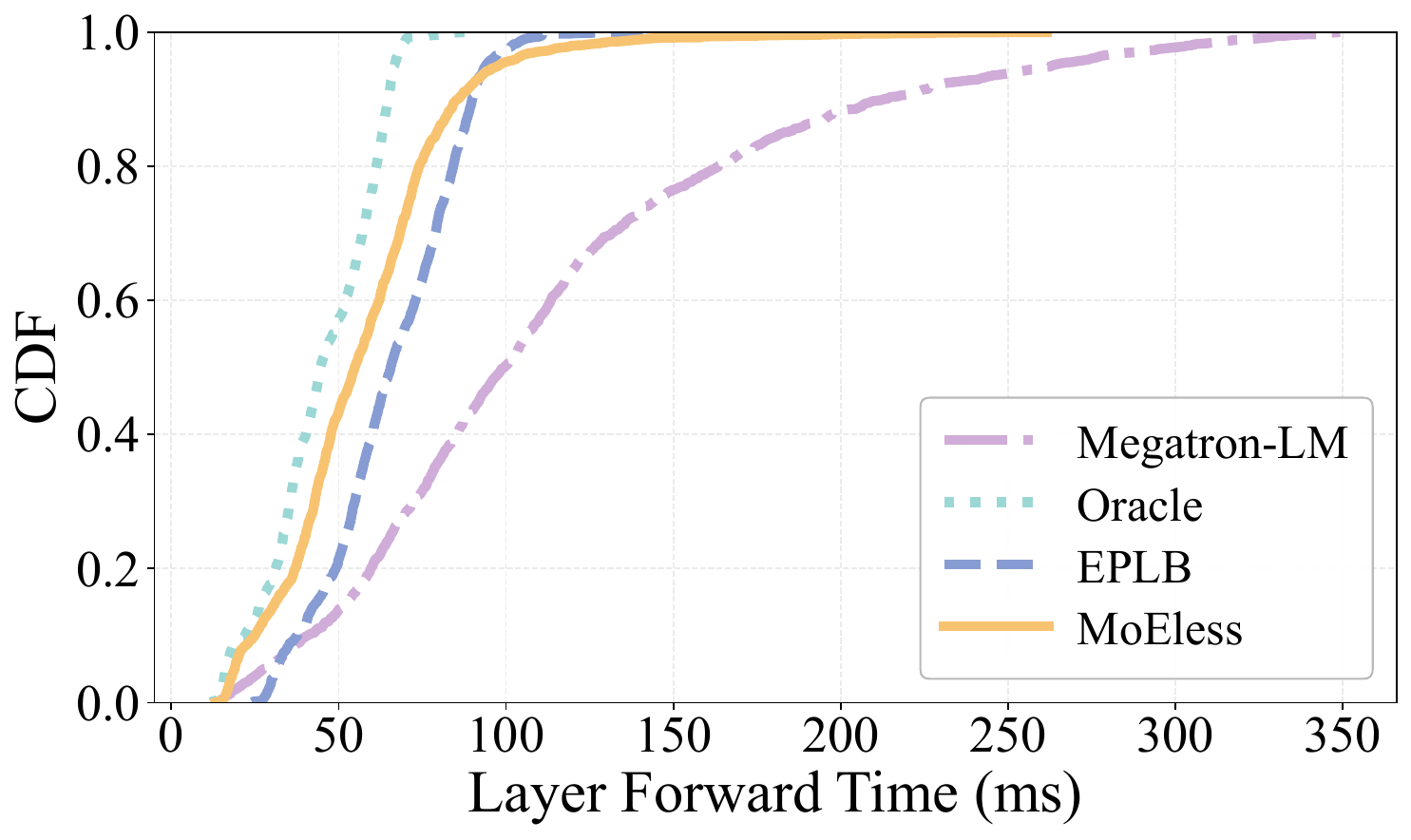}
        \caption{\llama.}
    \end{subfigure}
    \caption{MoE layer forward time of four approaches across three models on \sharegpt.}
    \label{fig:eval-layer-sharegpt}
    \vspace{6pt}
\end{figure*}

\begin{figure*}[t]
    \centering
    \begin{subfigure}[t]{0.46\linewidth}
        \centering
        \includegraphics[width=\linewidth]{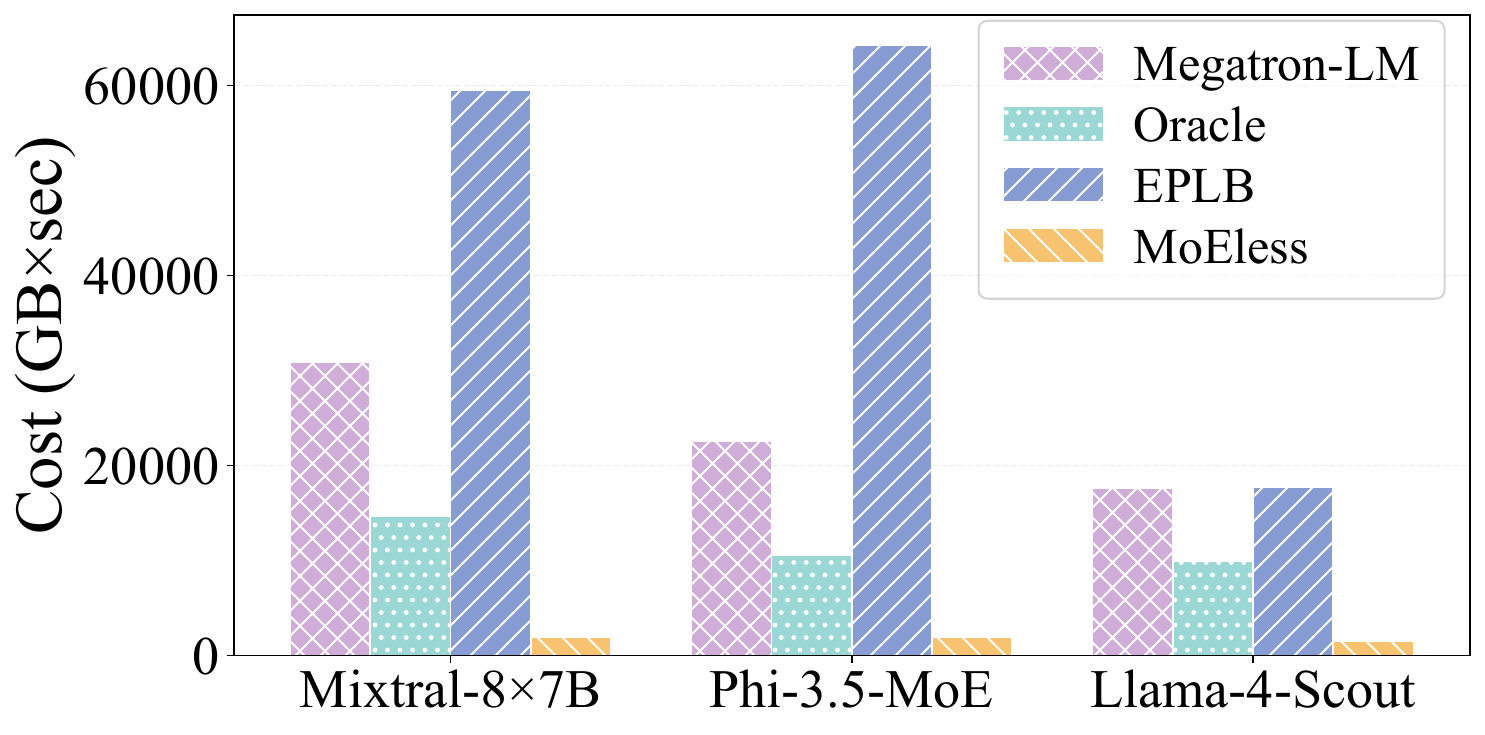}
        \caption{\lmsys.}
    \end{subfigure}
    \hfill
    \begin{subfigure}[t]{0.46\linewidth}
        \centering
        \includegraphics[width=\linewidth]{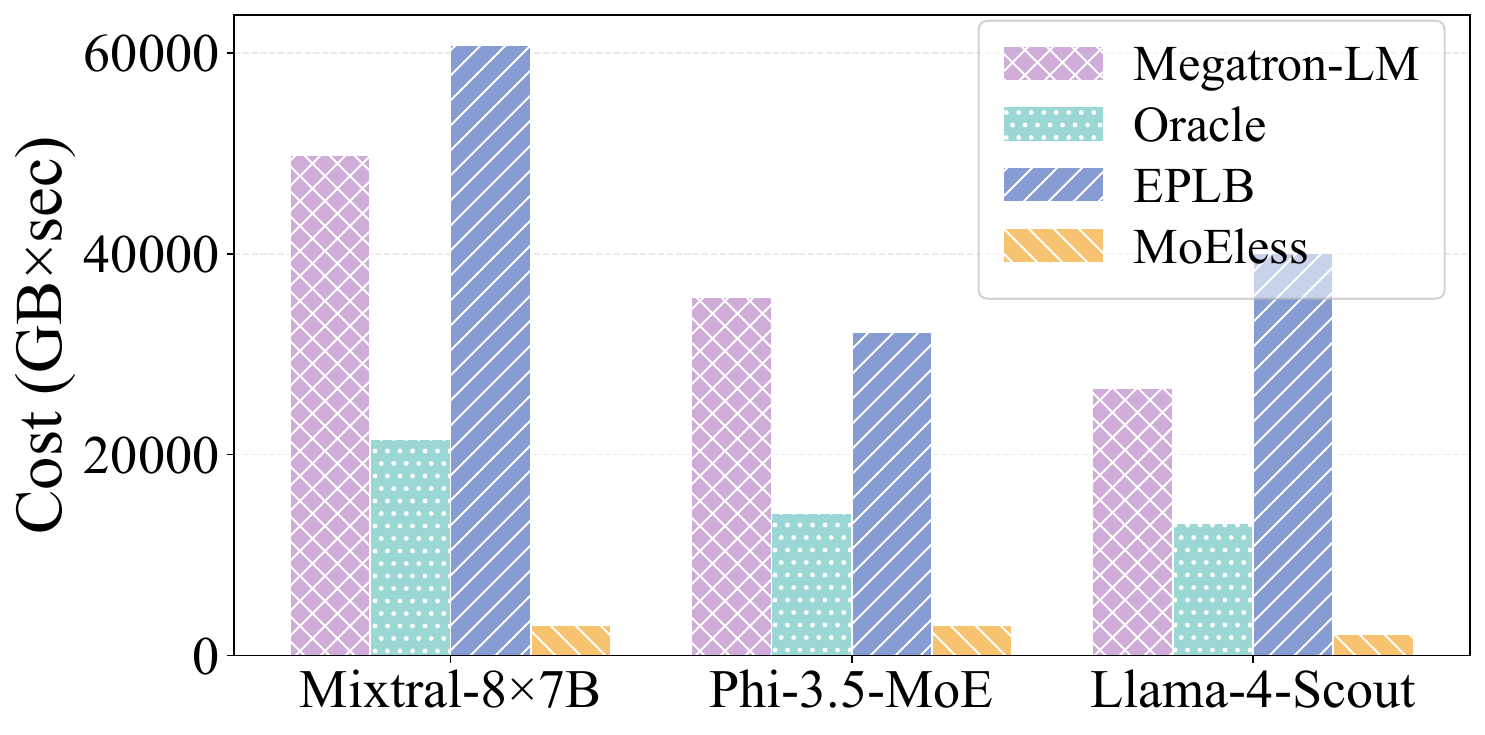}
        \caption{\sharegpt.}
    \end{subfigure}
    \caption{Total inference cost of four approaches across three models on two datasets.}
    \label{fig:eval-cost}
    \vspace{6pt}
\end{figure*}

\begin{figure*}[t]
    \centering
    \includegraphics[width=\linewidth]{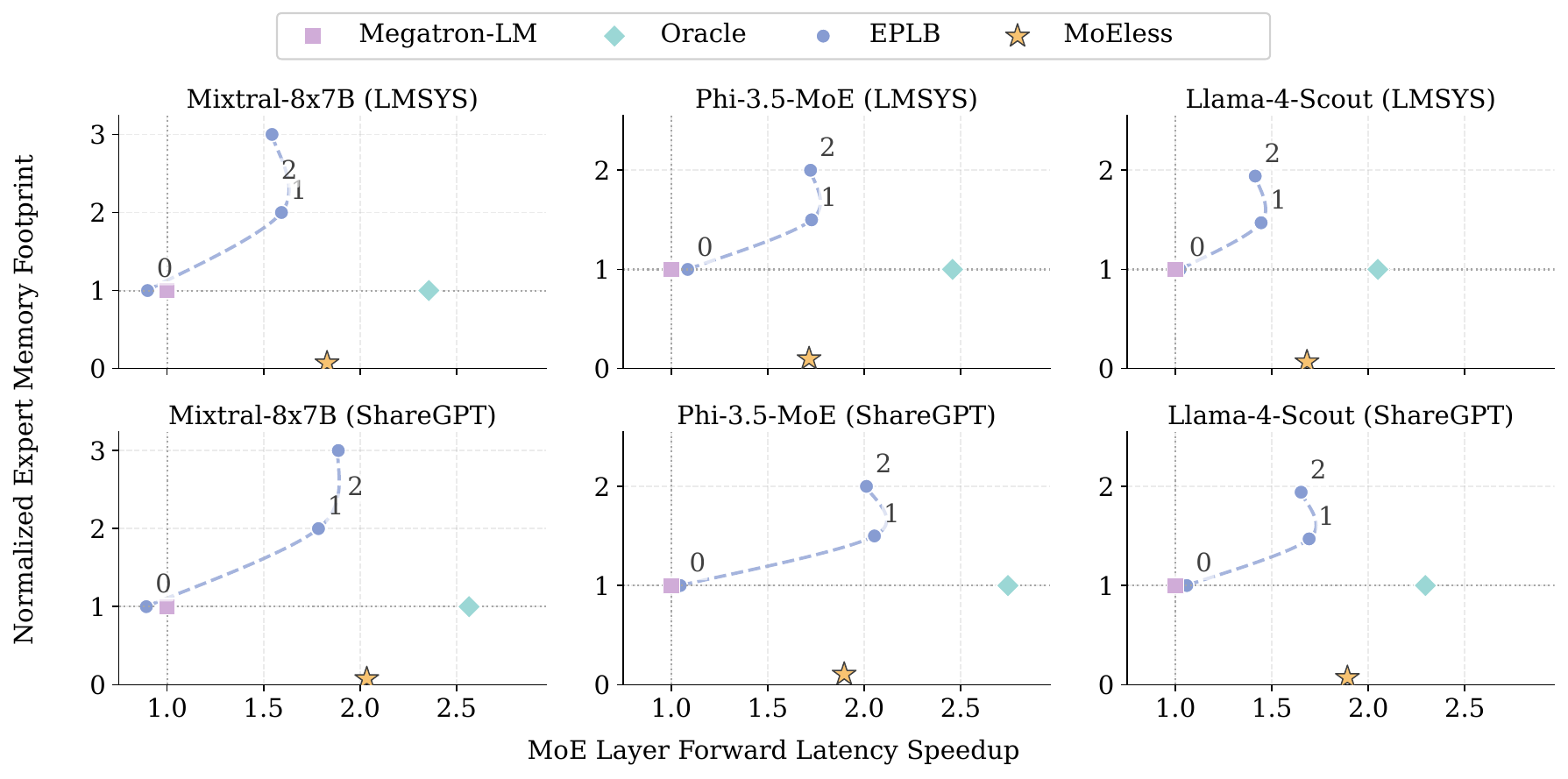}
    \caption{Performance tradeoff of \MoE layer forward latency speedup \vs expert memory cost.}
    \label{fig:eval-overall-tradeoff}
    \vspace{6pt}
\end{figure*}

\begin{figure}[t]
    \centering
    \includegraphics[width=\linewidth]{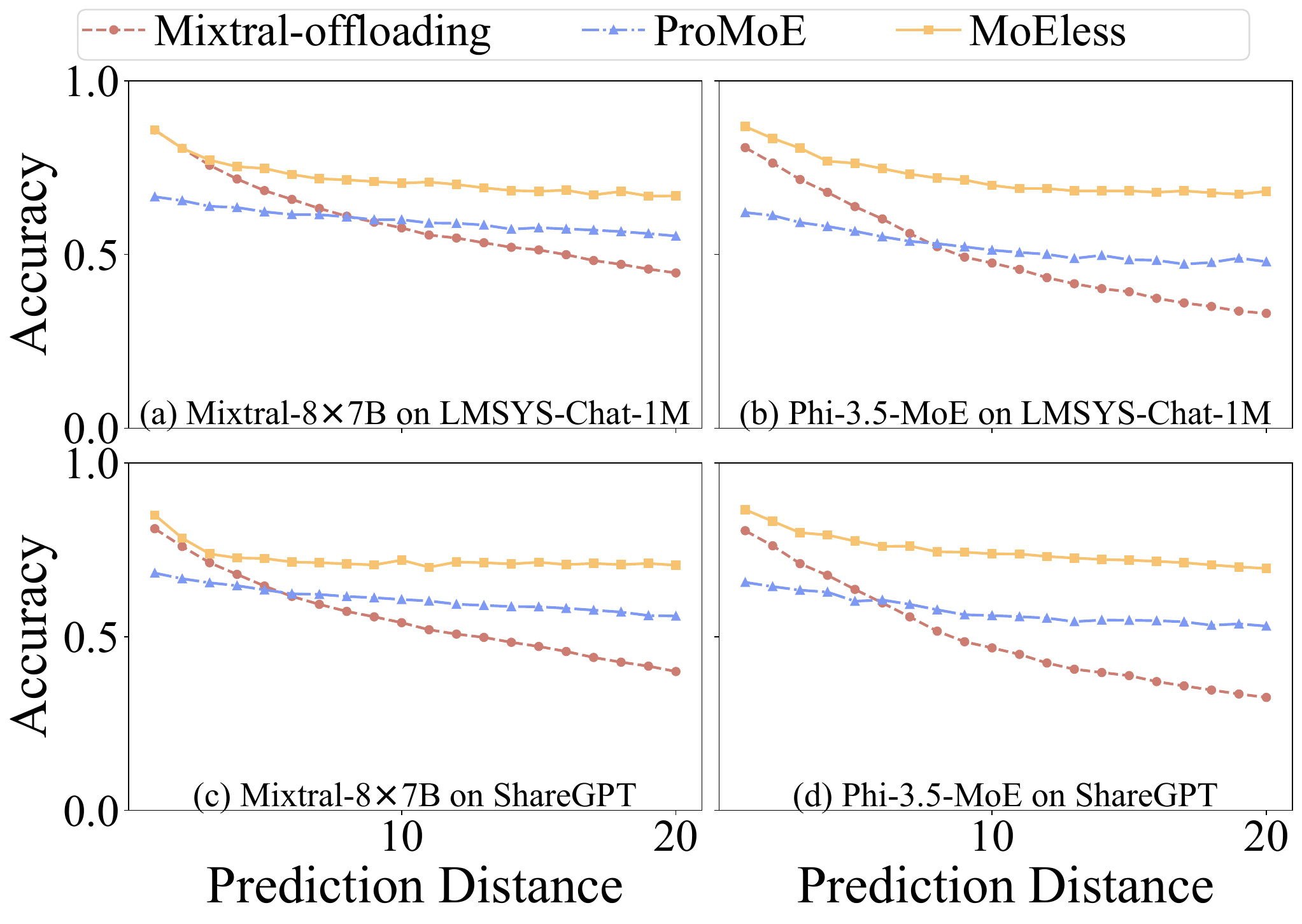}
    \caption{Expert load prediction accuracy for three prediction methods at different prediction distances.}
    \label{fig:PredictionAccuracy}
\end{figure}

\begin{figure}[t]
    \centering
    \begin{subfigure}[t]{0.49\linewidth}
        \centering
        \includegraphics[width=\linewidth]{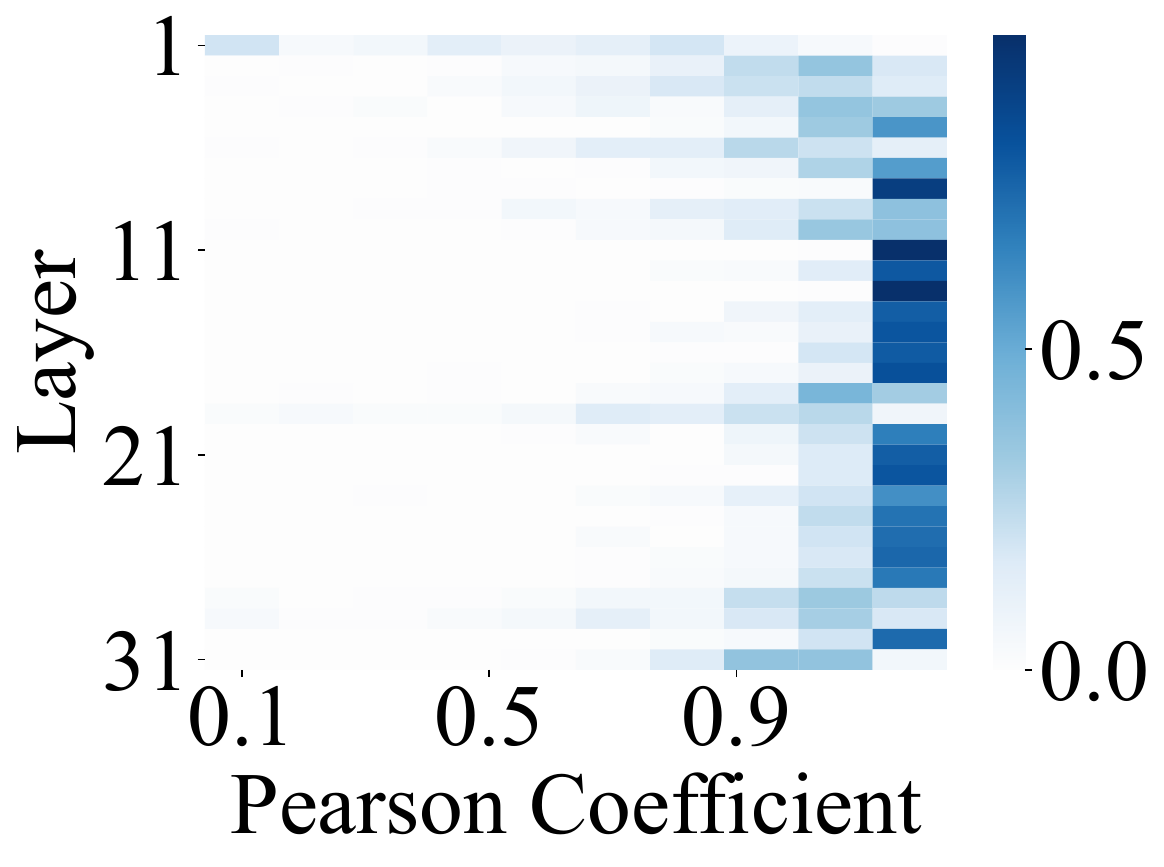}
        \caption{Mixtral-8x7B}
        \label{fig:mixtral_workload_coe_heatmap}
    \end{subfigure}
    \hfill
    \begin{subfigure}[t]{0.49\linewidth}
        \centering
        \includegraphics[width=\linewidth]{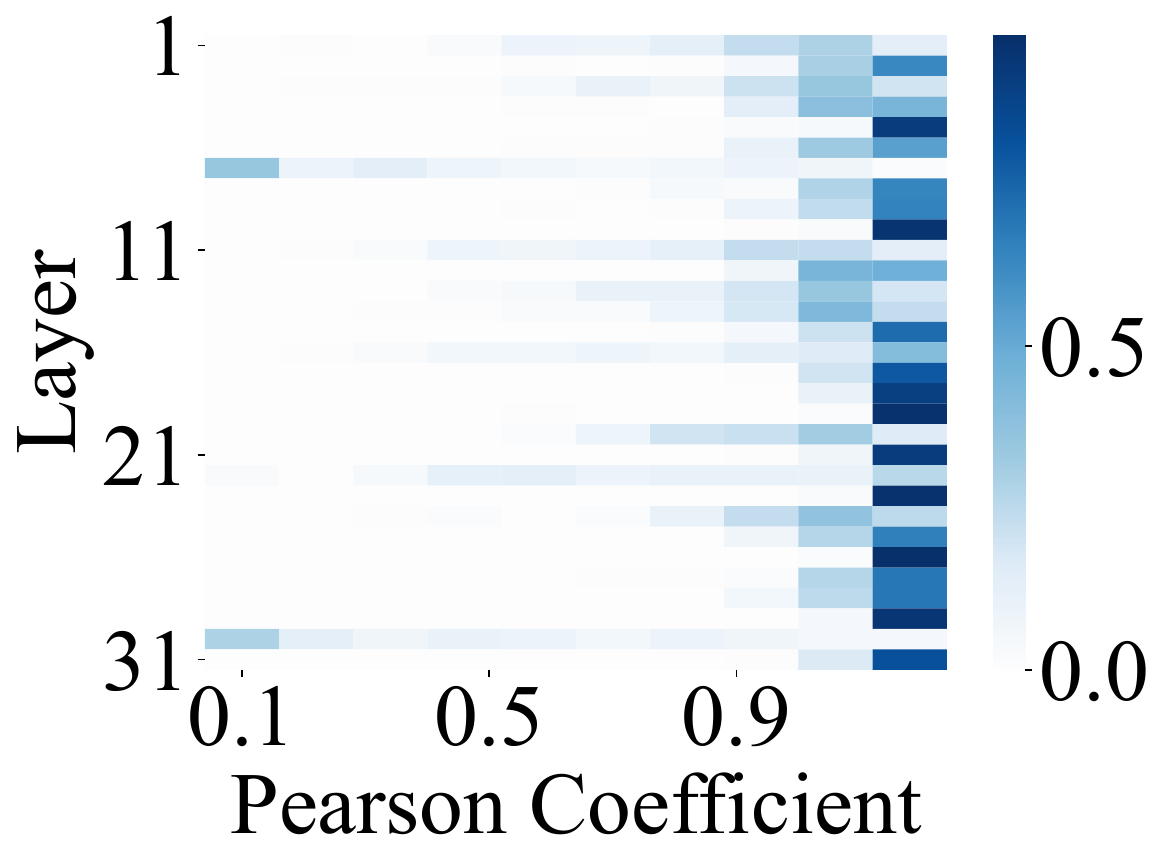}
        \caption{Phi-3.5-MoE}
        \label{fig:phi_workload_coe_heatmap}
    \end{subfigure}
    \caption{Correlations between predicted and actual expert load distributions across layers of two models. Heavier color means more correlation results fall in the slot.}  
    \label{fig:workload_coe}
\end{figure}

\textbf{Traces.}
We adopt real-world LLM inference traces released by Microsoft Azure~\cite{patel2024splitwise}.
As shown in Figure~\ref{fig:eval-azure-trace}, the traces exhibit bursty request arrivals and diverse input/output token lengths, capturing both temporal and computational variability in production \LLM serving.
We replay the traces by randomly sampling requests from the datasets and issuing them at their original timestamps.
Since \megatron does not natively support continuous batching, we aggregate all requests arriving within each second into a single input batch, resulting in time-varying batches that better reflect real-world serving scenarios.
Following prior work~\cite{patel2024splitwise,stojkovic2025dynamollm,zhong2024distserve,agrawal2024taming,yu2025taming}, each request processes and generates the exact numbers of input and output tokens specified by the traces.

\textbf{Testbed.} 
We conduct all experiments on an eight-GPU testbed, with GPUs interconnected via pairwise NVLinks. 
Each GPU is an NVIDIA A6000 equipped with 48 GB of GPU memory and connected to the host CPUs through PCIe 5.0 links, providing up to 64 GB/s bidirectional bandwidth per GPU. 
The system is provisioned with 256 vCPUs backed by AMD EPYC 7C13 processors and 500 GB of system memory. 

\textbf{Baselines.}
We compare \sys against three \SOTA serverful \MoE serving baselines:
1) \textbf{\megatron}~\cite{shoeybi2019megatron}: a basic \EP-enabled \MoE inference system without expert load balancing, and
2) \textbf{\EPLB} \cite{liu2024deepseek}: the load balancer proposed by DeepSeek \cite{liu2024deepseek}, which periodically (\eg, every ten minutes) creates redundant experts based on historical expert usage to mitigate high expert loads.
3) \textbf{Oracle}~\cite{he2026capacity}: an upper-bound baseline that ignores gate network outputs and performs perfect expert load balancing.
Note that Oracle directly affects model generation quality, as it ignores a subset of the original expert routing decisions during computation.
We integrate \EPLB and Oracle into \megatron for fair comparison.

\textbf{Metrics.}
We measure the \MoE layer forward latency to directly evaluate the overheads caused by expert stragglers.
We also report the overall serving cost, estimated as the product of memory consumption and inference latency aggregated across all input batches, representing the total monetary cost of serving the entire workload.
In addition, we evaluate detailed metrics, including expert load prediction accuracy and system overheads.

\begin{figure*}[t]
    \centering
    \begin{subfigure}[t]{0.328\linewidth}
        \centering
        \includegraphics[width=\linewidth]{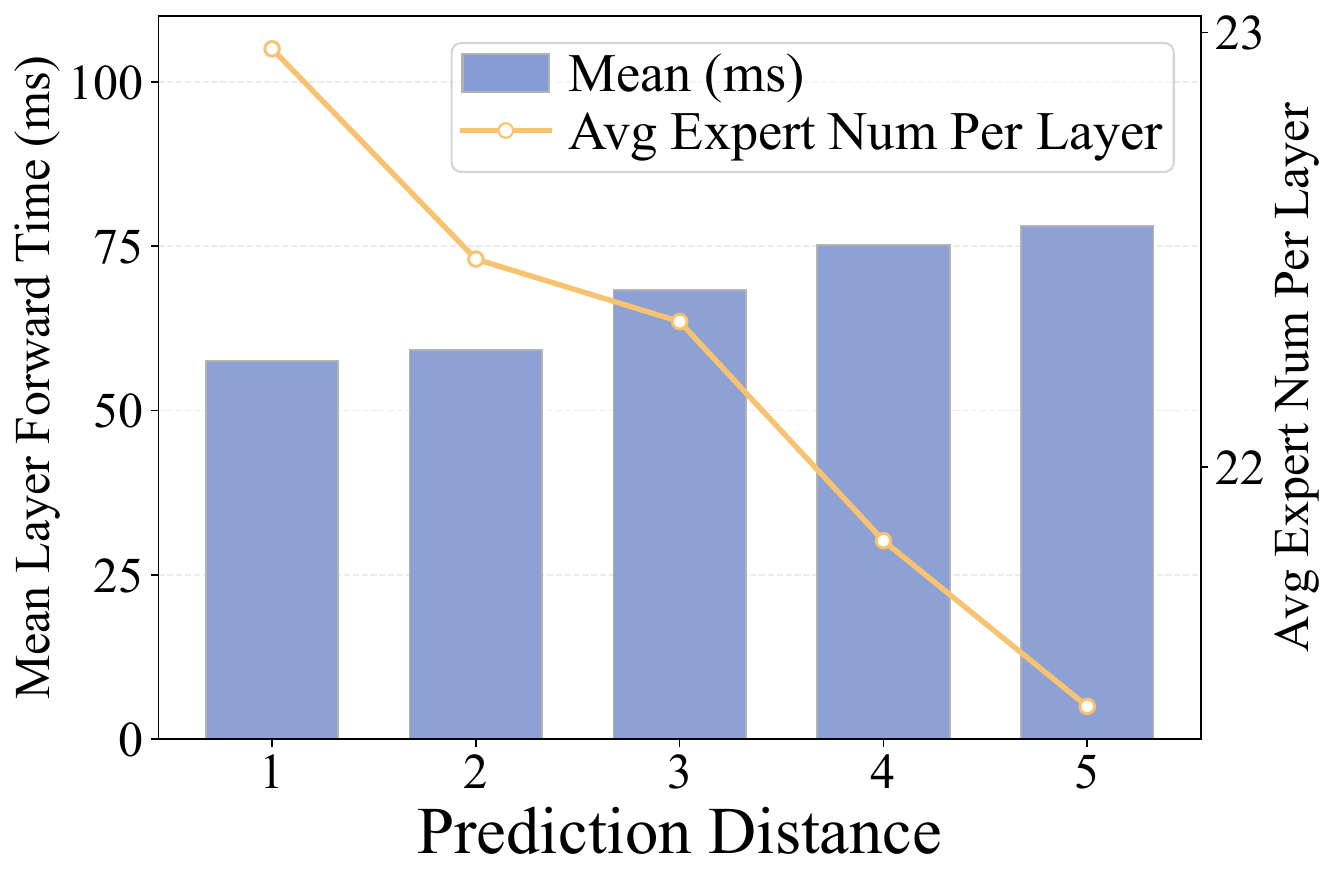}
        \caption{\mixtral.}
    \end{subfigure}
    \hfill
    \begin{subfigure}[t]{0.328\linewidth}
        \centering
        \includegraphics[width=\linewidth]{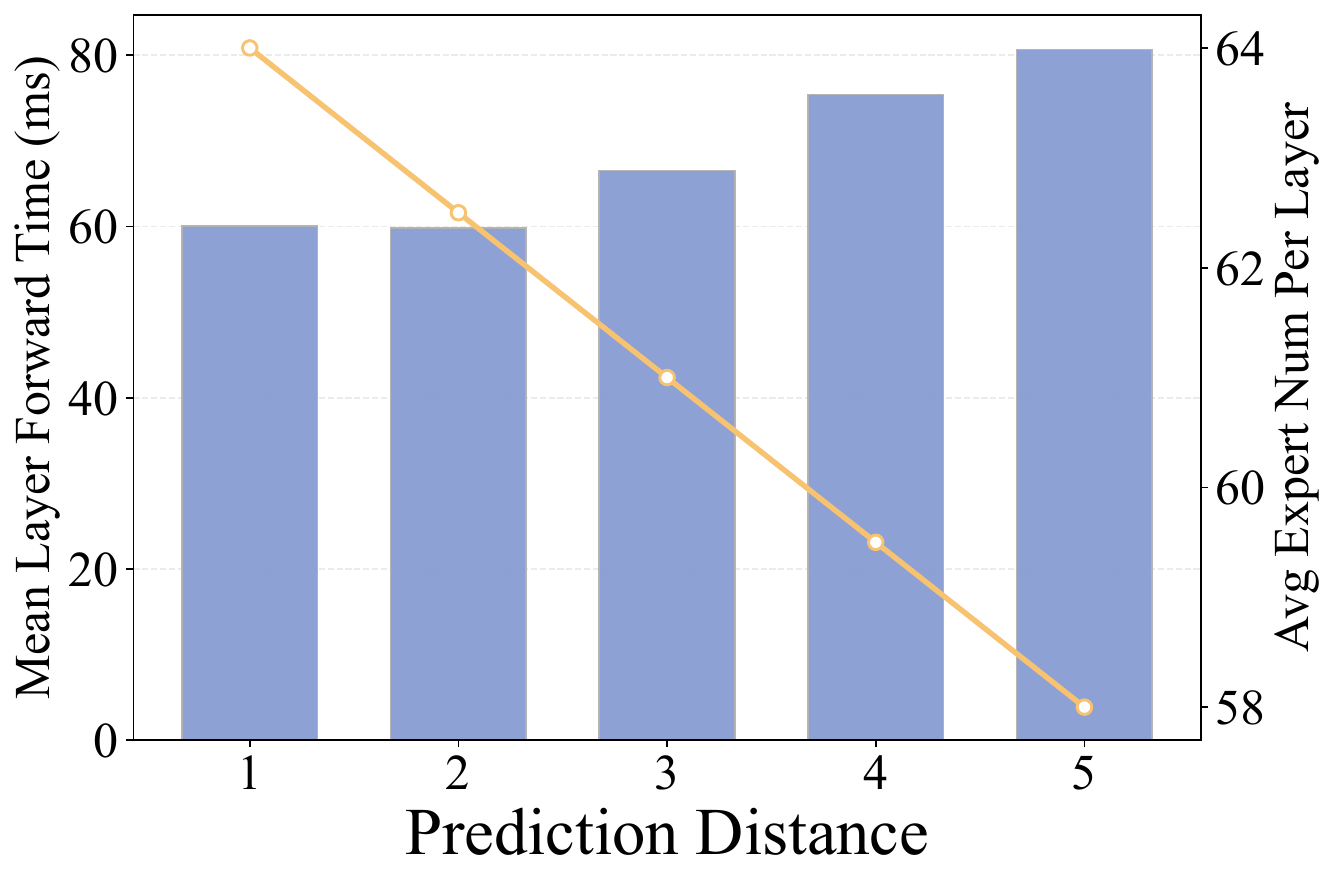}
        \caption{\phimoe.}
    \end{subfigure}
    \hfill
    \begin{subfigure}[t]{0.328\linewidth}
        \centering
        \includegraphics[width=\linewidth]{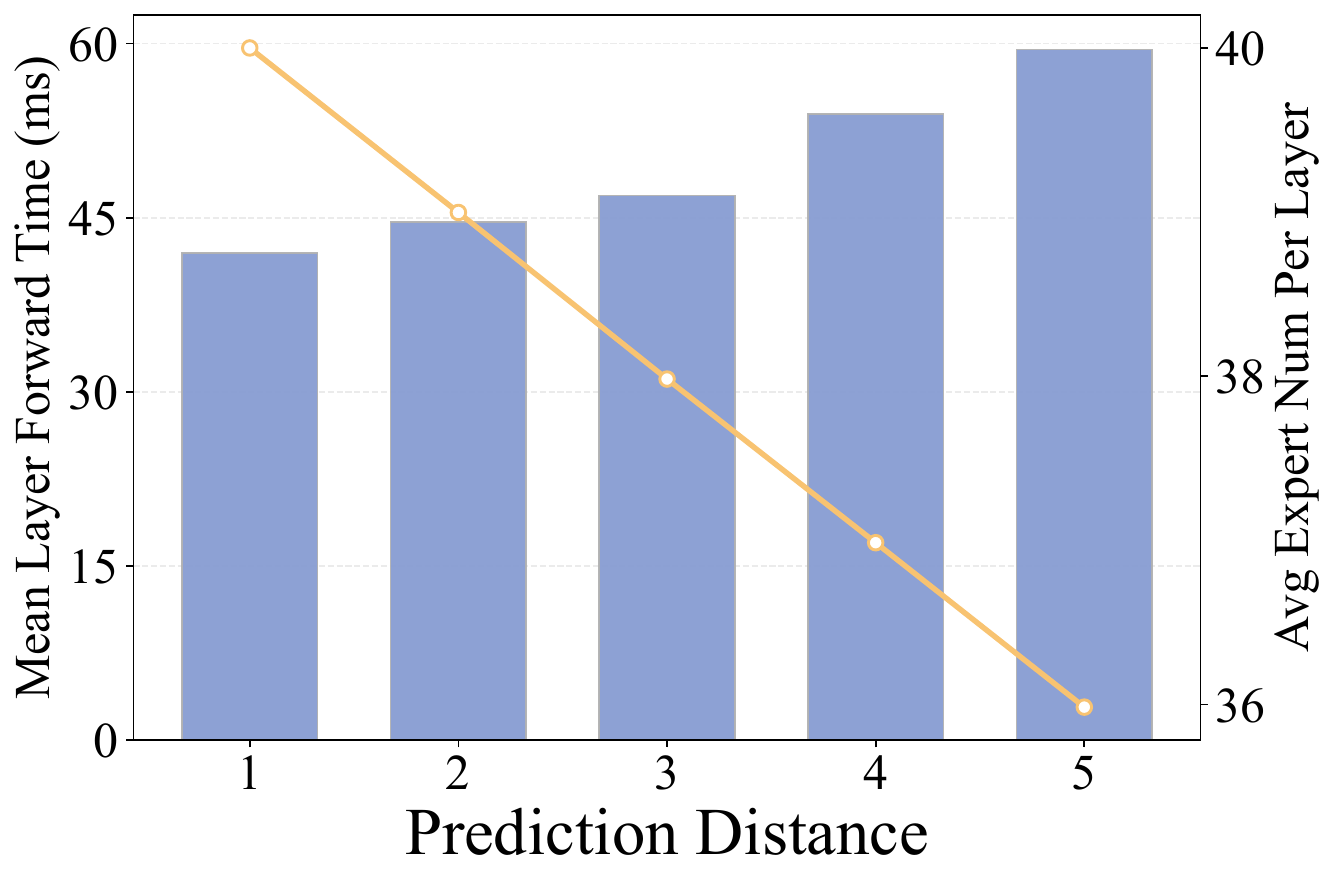}
        \caption{\llama.}
    \end{subfigure}
    \caption{Sensitivity analysis of \sys's expert prediction distance on \lmsys.}
    \label{fig:eval-sensitivity-predict-lmsys}
    \vspace{6pt}
\end{figure*}

\begin{figure*}[t]
    \centering
    \begin{subfigure}[t]{0.328\linewidth}
        \centering
        \includegraphics[width=\linewidth]{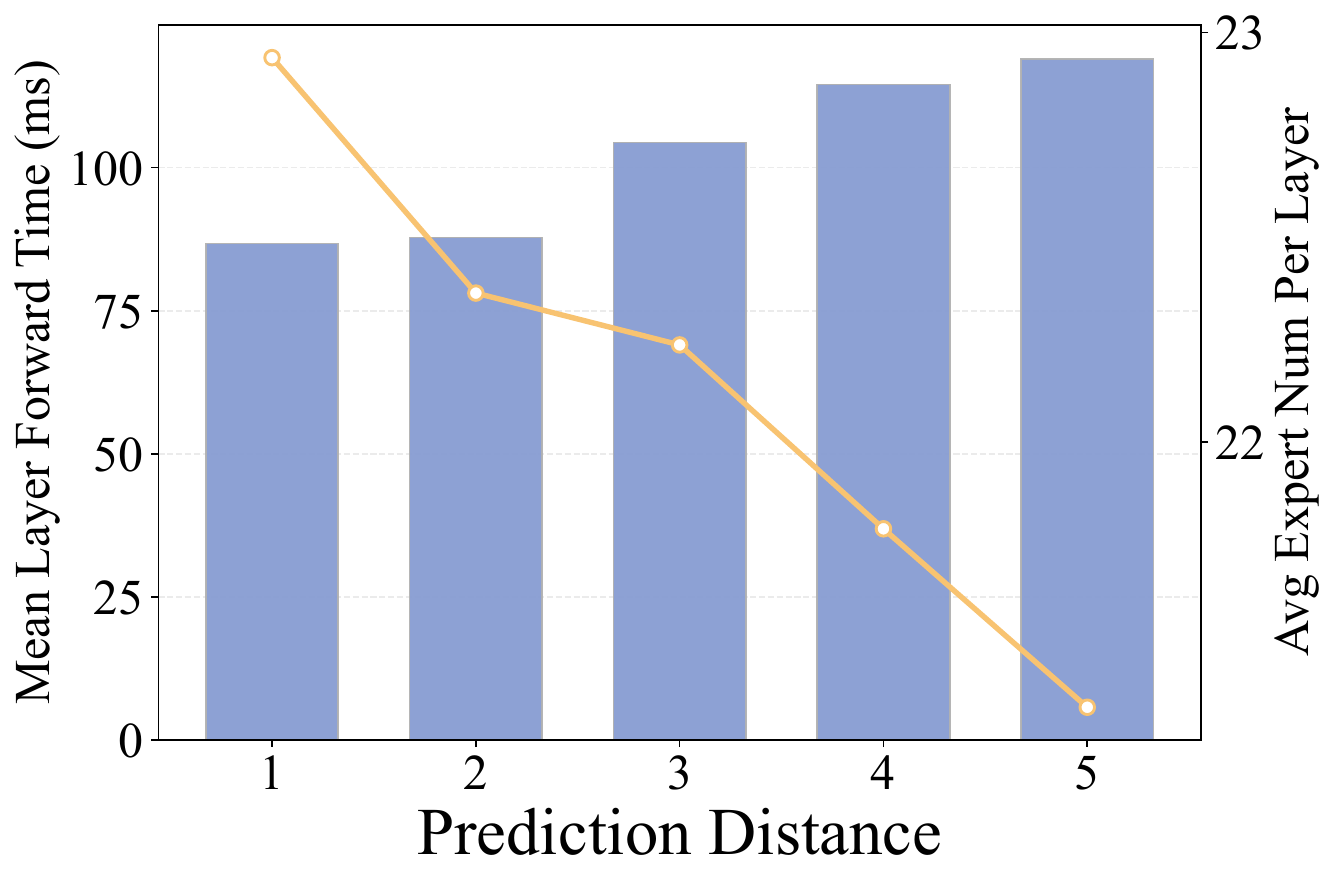}
        \caption{\mixtral.}
    \end{subfigure}
    \hfill
    \begin{subfigure}[t]{0.328\linewidth}
        \centering
        \includegraphics[width=\linewidth]{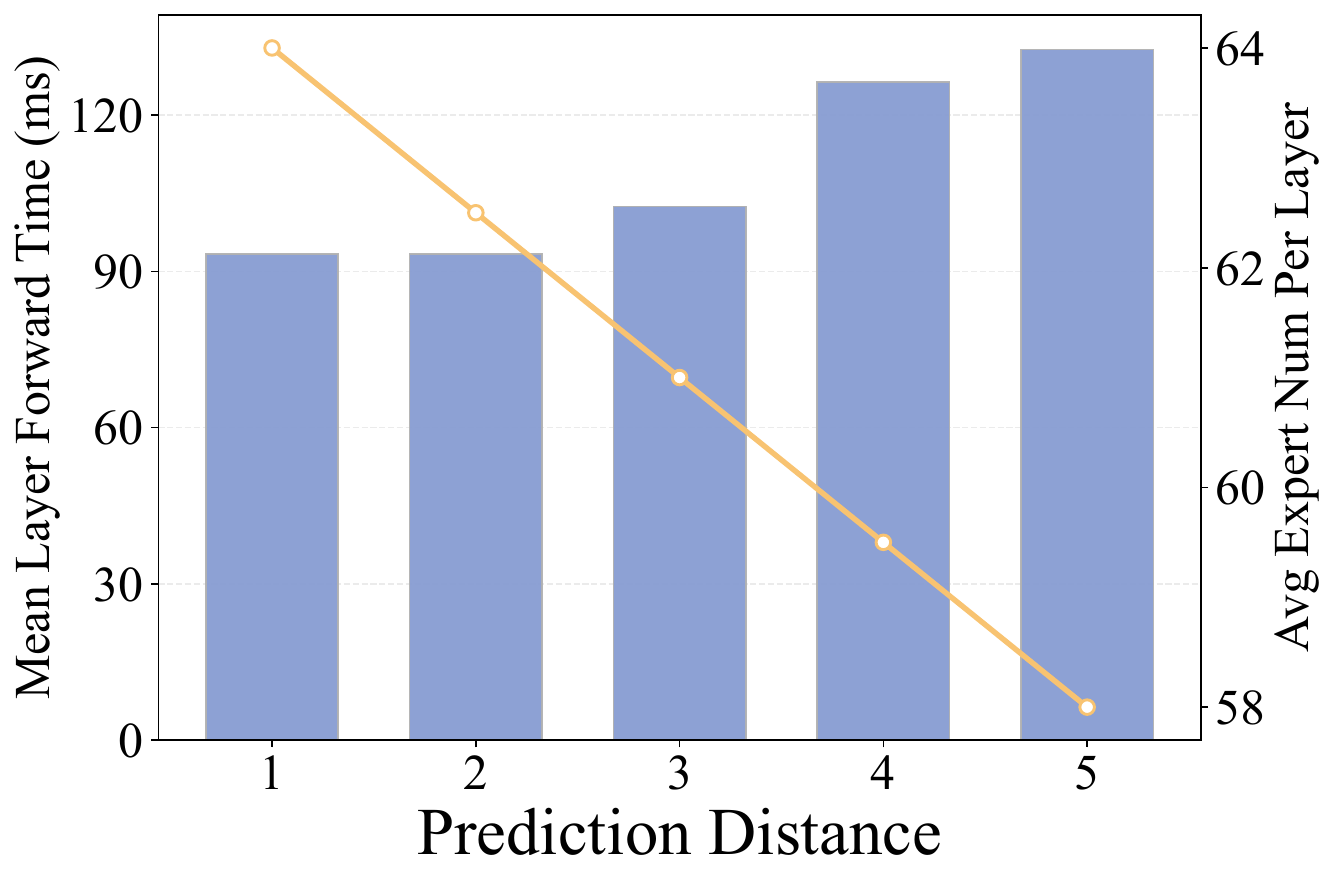}
        \caption{\phimoe.}
    \end{subfigure}
    \hfill
    \begin{subfigure}[t]{0.328\linewidth}
        \centering
        \includegraphics[width=\linewidth]{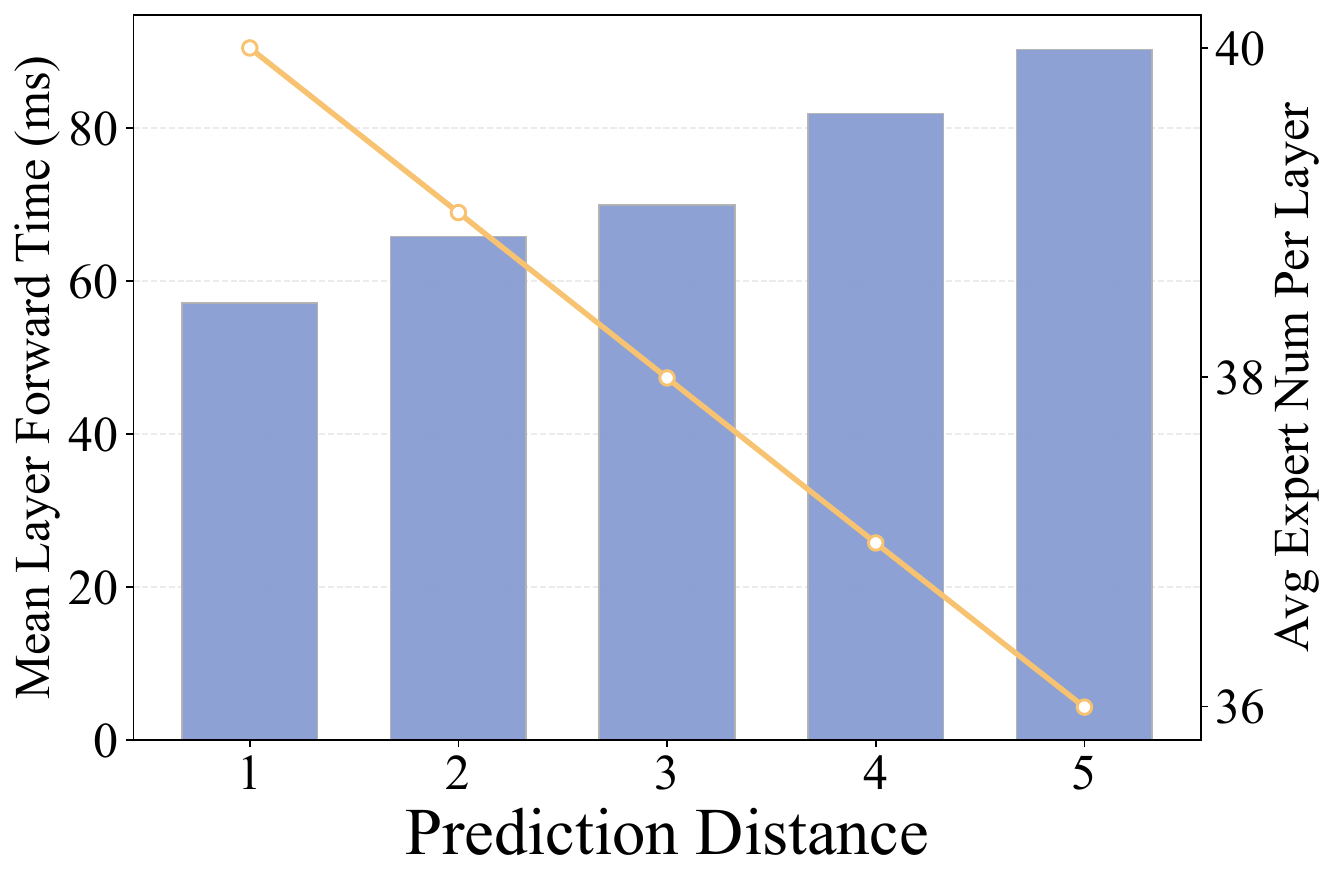}
        \caption{\llama.}
    \end{subfigure}
    \caption{Sensitivity analysis of \sys's expert prediction distance on \sharegpt.}
    \label{fig:eval-sensitivity-predict-sharegpt}
    \vspace{6pt}
\end{figure*}

\subsection{Overall Performance}
\label{subsec:eval-overall}

We evaluate the overall performance of \megatron, Oracle, \EPLB, and \sys by running three \MoE models on the \lmsys and \sharegpt datasets, measuring \MoE layer forward latency and total inference cost.
Due to continuous batching, batch-level latency varies across iterations.
Therefore, we record the per-layer forward latency for all layers across all input batches and aggregate them into a unified \CDF for comparison.

Figures~\ref{fig:eval-layer-lmsys} and \ref{fig:eval-layer-sharegpt} present the \CDF of \MoE layer forward latency of four approaches across three models on \lmsys and \sharegpt, respectively.
With scalable and elastic serverless experts, \sys achieves significant performance improvements, reducing the average \MoE layer forward latency by 43.19\% and 21.89\% compared to \megatron and \EPLB, respectively.
While Oracle is a lossy baseline that achieves perfect load balance by affecting generation quality, \sys consistently stays closest to Oracle across all cases, indicating superior performance over existing \SOTA expert load balancing methods.

Figure~\ref{fig:eval-cost} reports the total inference cost aggregated across the entire experiment for all four approaches.
All other baselines incur significantly higher costs due to serverful expert execution.
In contrast, by leveraging serverless expert execution, \sys consistently delivers higher serving efficiency, reducing overall inference cost by 92.68\%, 84.06\%, and 95.11\% compared to \megatron, Oracle, and \EPLB, respectively.

Figure~\ref{fig:eval-overall-tradeoff} compares the latency--memory tradeoff of all four approaches by showing normalized expert memory footprint against \MoE layer forward latency speedup.
Across all models and datasets, \EPLB exhibits the expected latency--memory tradeoff. As the number of expert replicas increases (from 0 to 2 for each expert), more expert replicas are provisioned to alleviate load imbalance, resulting in larger memory footprints. 
However, beyond a certain point, the additional expert replicas introduce higher communication and synchronization overhead, so the most aggressive configuration does not consistently deliver the lowest latency.
Oracle provides the upper bound in latency improvement due to lossy expert routing while still maintaining the full expert memory footprint.
In contrast, \sys consistently occupies the Pareto-optimal region, achieving latency close to Oracle while reducing the expert memory footprint by more than one order of magnitude relative to all serverful baselines.
This demonstrates that elastic serverless experts enable both efficient expert utilization and high serving performance without sacrificing memory efficiency.

\subsection{Prediction Accuracy}
\label{subsec:eval-prediction}

We compare \sys's predictor against two \SOTA expert prediction solutions:
1) \textbf{Mixtral-offloading}~\cite{eliseev2023fast}, which directly employs the original gate networks to predict expert selections of subsequent layers, and
2) \textbf{ProMoE}~\cite{song2024promoe}, which trains a large, layer-specific predictor from scratch to model the mapping between gate inputs and expert selections.
We use the same experimental setup as in \S\ref{subsec:eval-overall}.

As shown in Figure~\ref{fig:PredictionAccuracy}, \sys consistently outperforms both baselines across different prediction distances, demonstrating greater robustness and accuracy.
On average, \sys improves prediction accuracy by up to 18\% and 15\% over Mixtral-offloading and ProMoE, respectively.

Figure~\ref{fig:workload_coe} presents the heatmap comparing the predicted and actual expert load distributions.
We collect pairwise predicted–actual data points across all layers and compute the Pearson correlation coefficients~\cite{cohen2009pearson} for each pair.
The results reveal a strong positive correlation between the predicted and actual distributions, indicating that our predictor effectively captures expert load patterns under real-world workloads.

\begin{figure*}[t]
    \centering
    \begin{subfigure}[t]{0.328\linewidth}
        \centering
        \includegraphics[width=\linewidth]{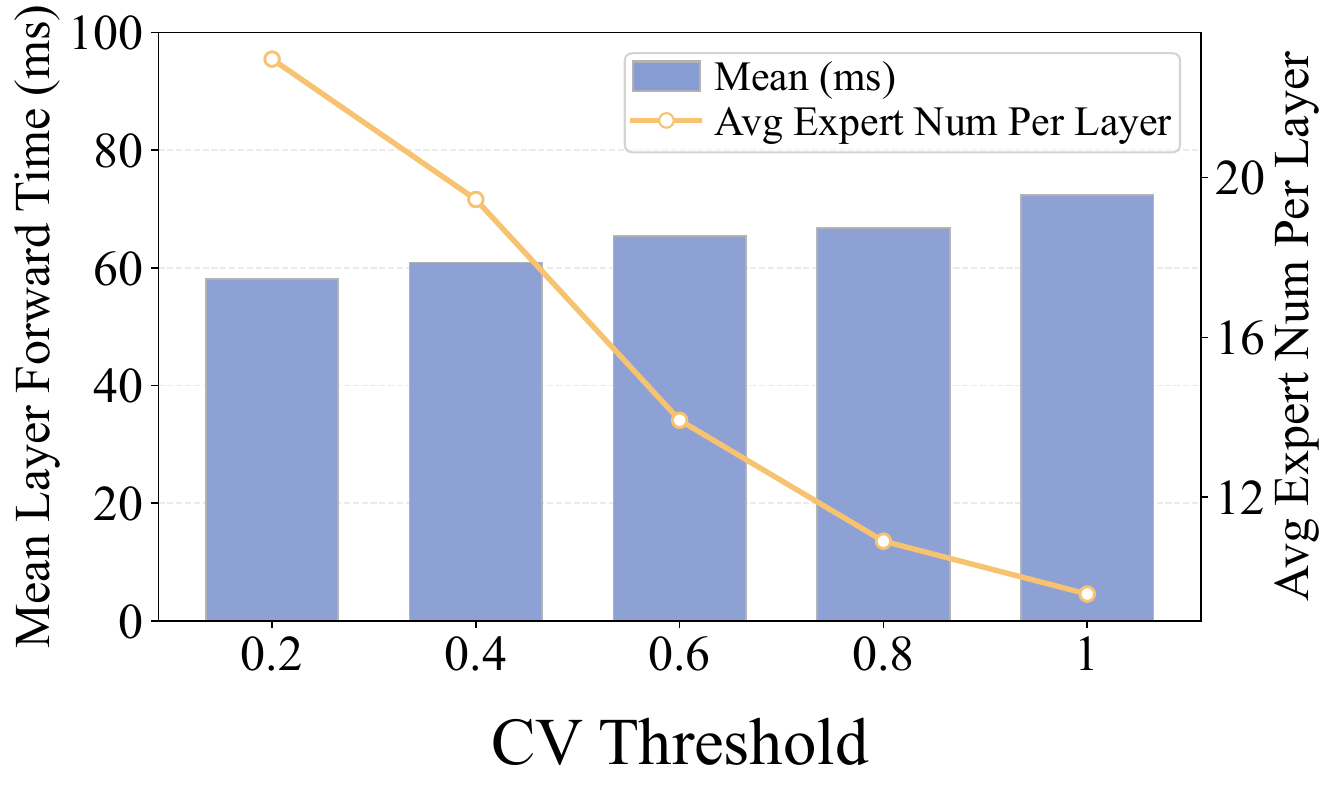}
        \caption{\mixtral.}
    \end{subfigure}
    \hfill
    \begin{subfigure}[t]{0.328\linewidth}
        \centering
        \includegraphics[width=\linewidth]{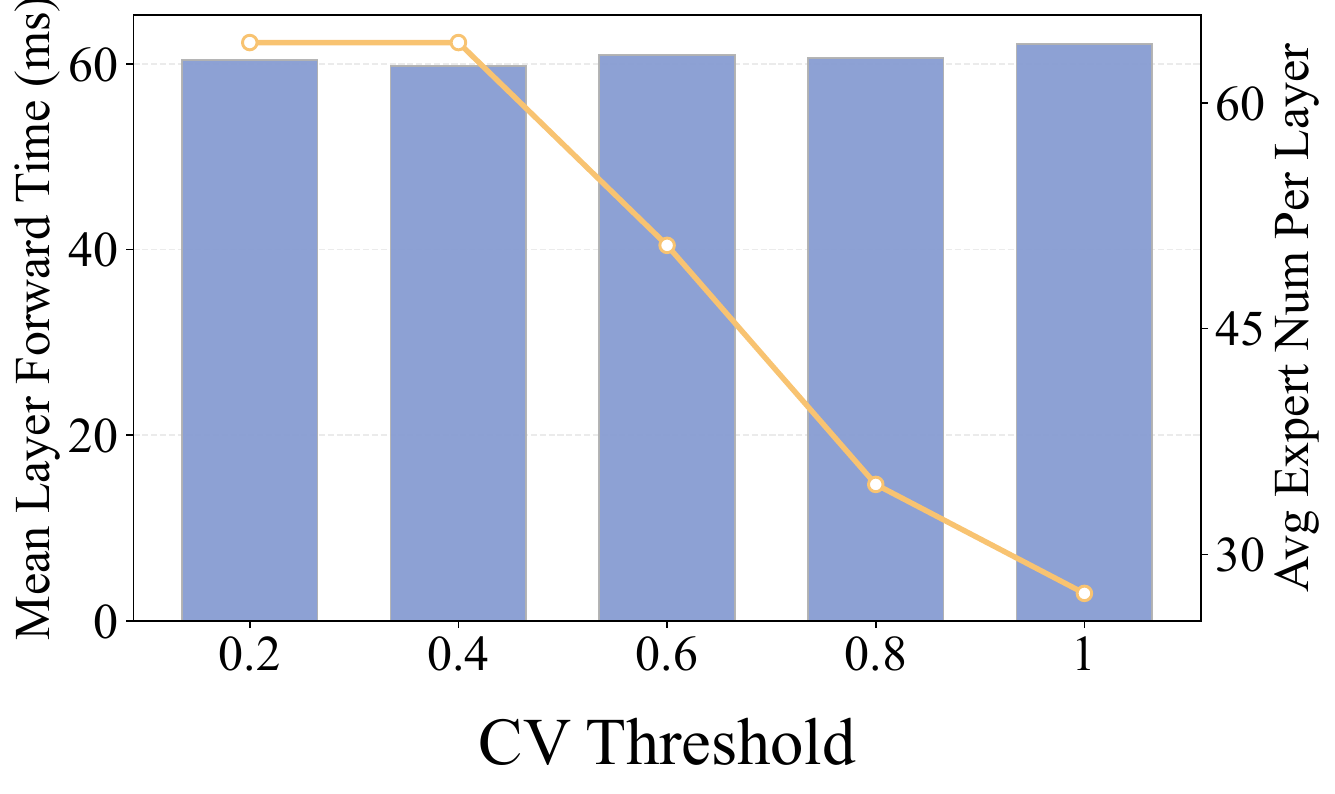}
        \caption{\phimoe.}
    \end{subfigure}
    \hfill
    \begin{subfigure}[t]{0.328\linewidth}
        \centering
        \includegraphics[width=\linewidth]{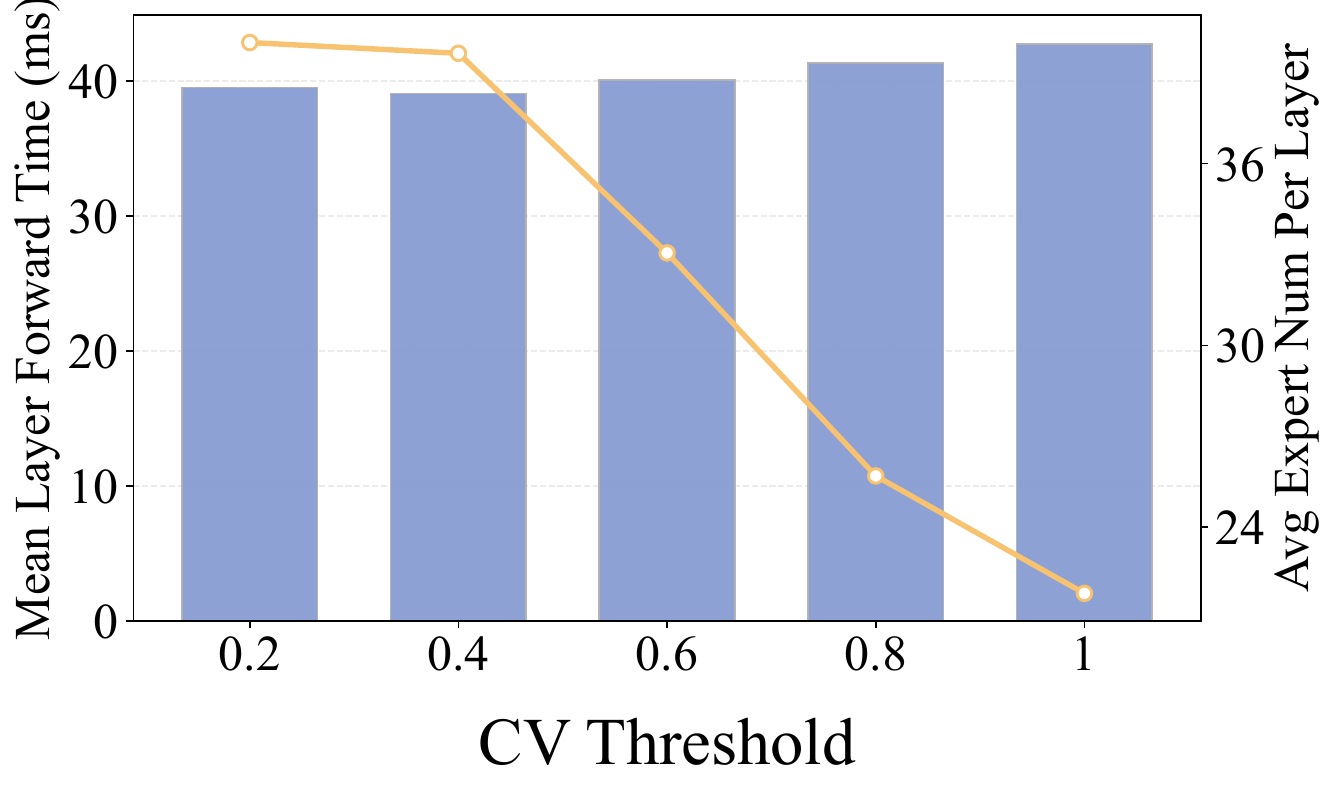}
        \caption{\llama.}
    \end{subfigure}
    \caption{Sensitivity analysis of \sys's expert load \CV threshold on \lmsys.}
    \label{fig:eval-sensitivity-cv-lmsys}
    \vspace{6pt}
\end{figure*}

\begin{figure*}[t]
    \centering
    \begin{subfigure}[t]{0.328\linewidth}
        \centering
        \includegraphics[width=\linewidth]{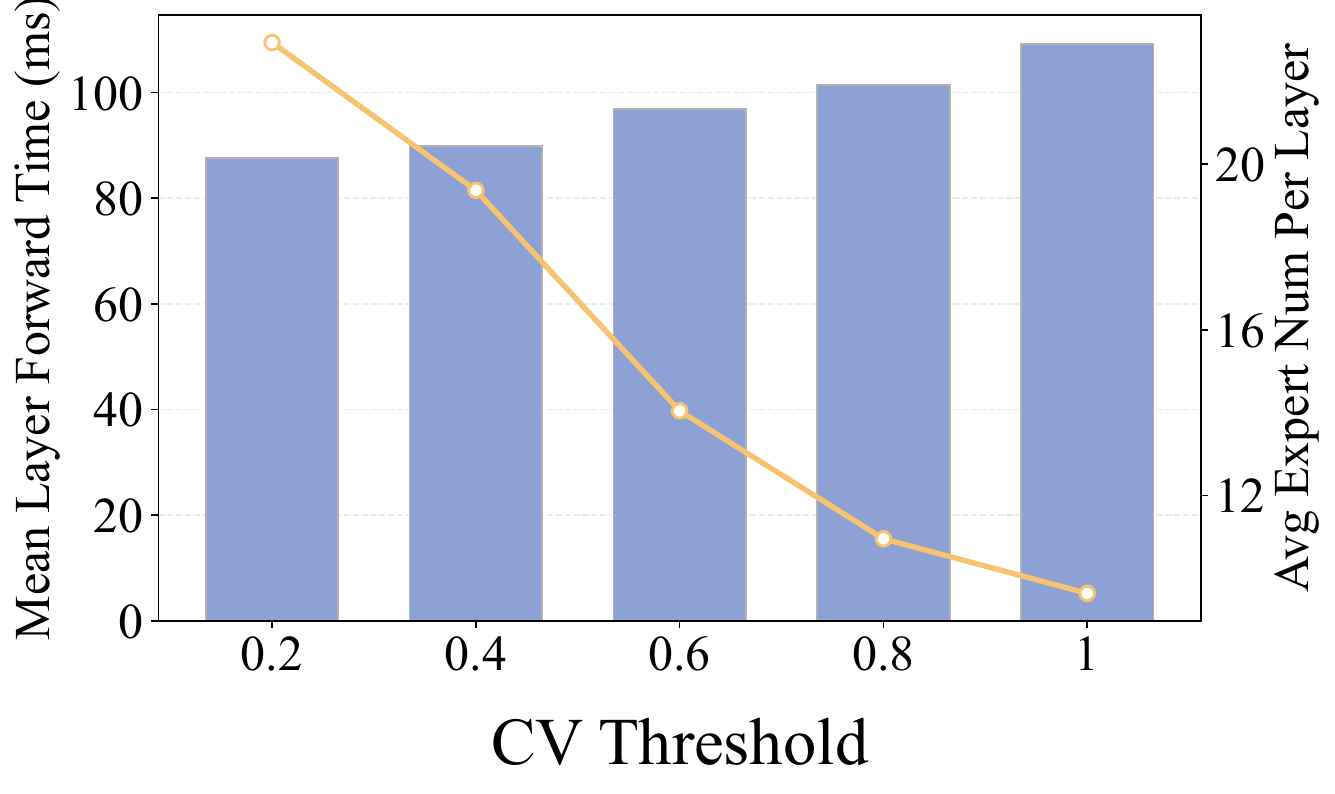}
        \caption{\mixtral.}
    \end{subfigure}
    \hfill
    \begin{subfigure}[t]{0.328\linewidth}
        \centering
        \includegraphics[width=\linewidth]{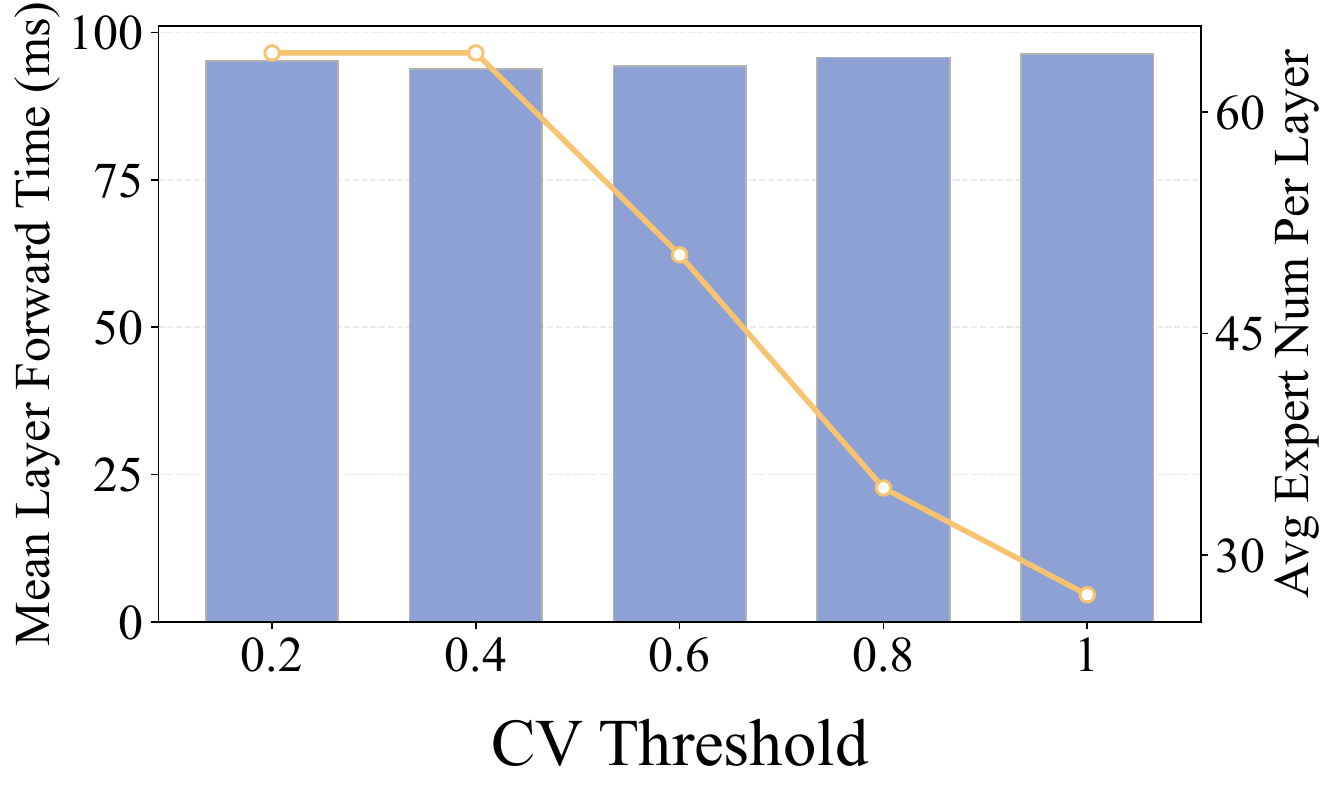}
        \caption{\phimoe.}
    \end{subfigure}
    \hfill
    \begin{subfigure}[t]{0.328\linewidth}
        \centering
        \includegraphics[width=\linewidth]{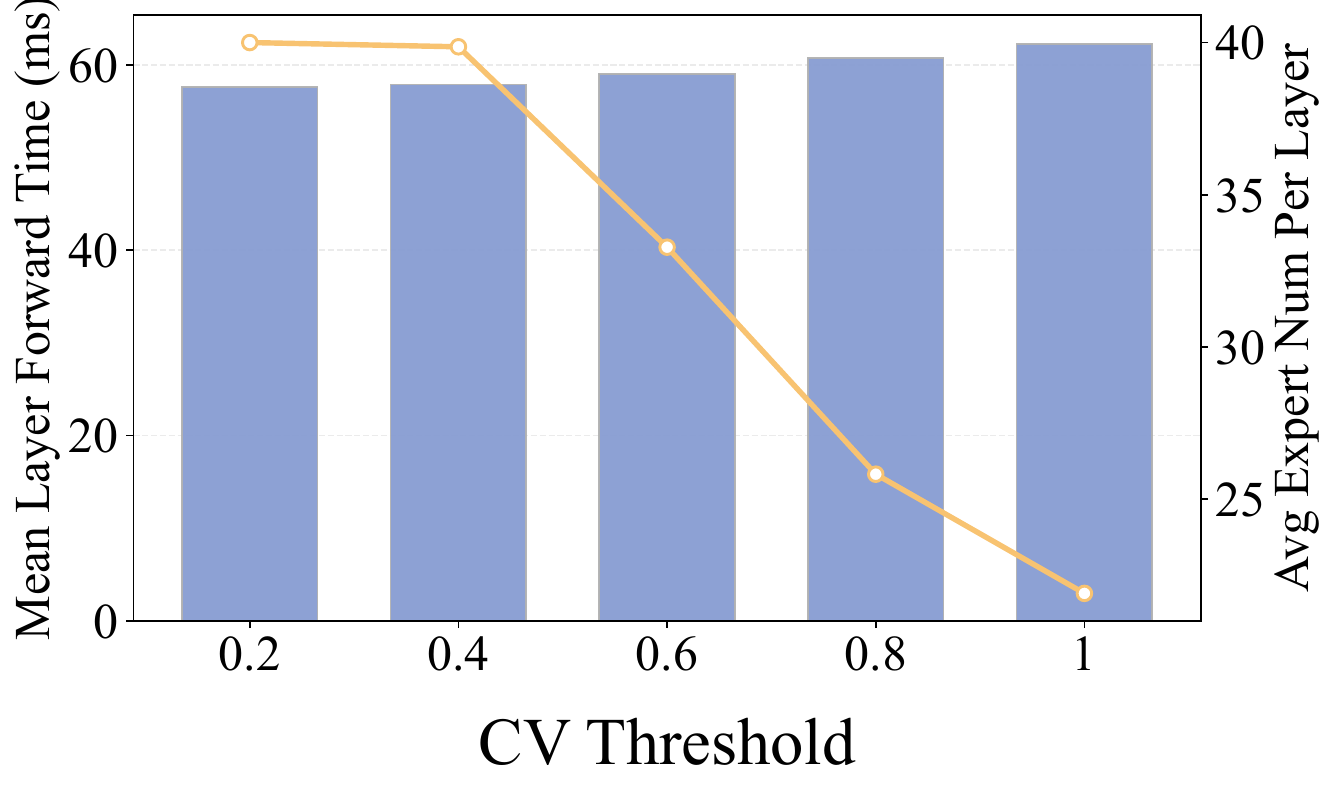}
        \caption{\llama.}
    \end{subfigure}
    \caption{Sensitivity analysis of \sys's expert load \CV threshold on \sharegpt.}
    \label{fig:eval-sensitivity-cv-sharegpt}
    \vspace{6pt}
\end{figure*}

\subsection{Sensitivity Analysis}
\label{subsec:eval-sensitivity}

We evaluate the sensitivity of two key system parameters in \sys: 1) the expert prediction distance, which determines the overlap overhead (\S\ref{subsec:design-predictor}), and 2) the expert load \CV threshold, which governs expert replica scaling (\S\ref{subsec:design-scaler}).
For both parameters, we measure the average \MoE layer forward time and the average number of expert replicas per layer to analyze their impact on inference latency and expert serving cost.

\textbf{Expert prediction distance.}
Figures~\ref{fig:eval-sensitivity-predict-lmsys} and~\ref{fig:eval-sensitivity-predict-sharegpt} show the average \MoE layer forward time and the average number of expert replicas per layer as the prediction distance varies, evaluated across three models and two datasets.
We increase the prediction distance from 1 to 5 in increments of 1. 
As the prediction distance increases, the \MoE layer forward time rises due to less accurate estimation of future expert load distributions, while the number of expert replicas decreases as load predictions become coarser.
Based on this trade-off, we set the prediction distance of \sys to 1 in all evaluations, achieving high prediction accuracy with negligible inference overhead.

\textbf{Expert load CV threshold.}
Figures~\ref{fig:eval-sensitivity-cv-lmsys} and~\ref{fig:eval-sensitivity-cv-sharegpt} present the sensitivity of the average \MoE layer forward time and the average number of expert replicas per layer to different \CV thresholds, evaluated across three models and two datasets.
We vary the \CV threshold from 0.2 to 1.0 in increments of 0.2. 
Larger \CV thresholds permit greater load imbalance and trigger less aggressive expert scaling, reducing the number of expert replicas per layer but increasing the \MoE layer forward time due to straggler effects.
We therefore set the \CV threshold of \sys to 0.2 in the evaluation.

\subsection{Ablation Study}
\label{subsec:eval-ablation}

We evaluate the contribution of each component in \sys by independently disabling the \textbf{Expert Load Predictor}, \textbf{Expert Scaler}, and \textbf{Expert Placer}, while keeping all remaining components unchanged.
We follow the same evaluation setup as in \S\ref{subsec:eval-overall}.

\begin{figure}[t]
    \centering
    \includegraphics[width=\linewidth]{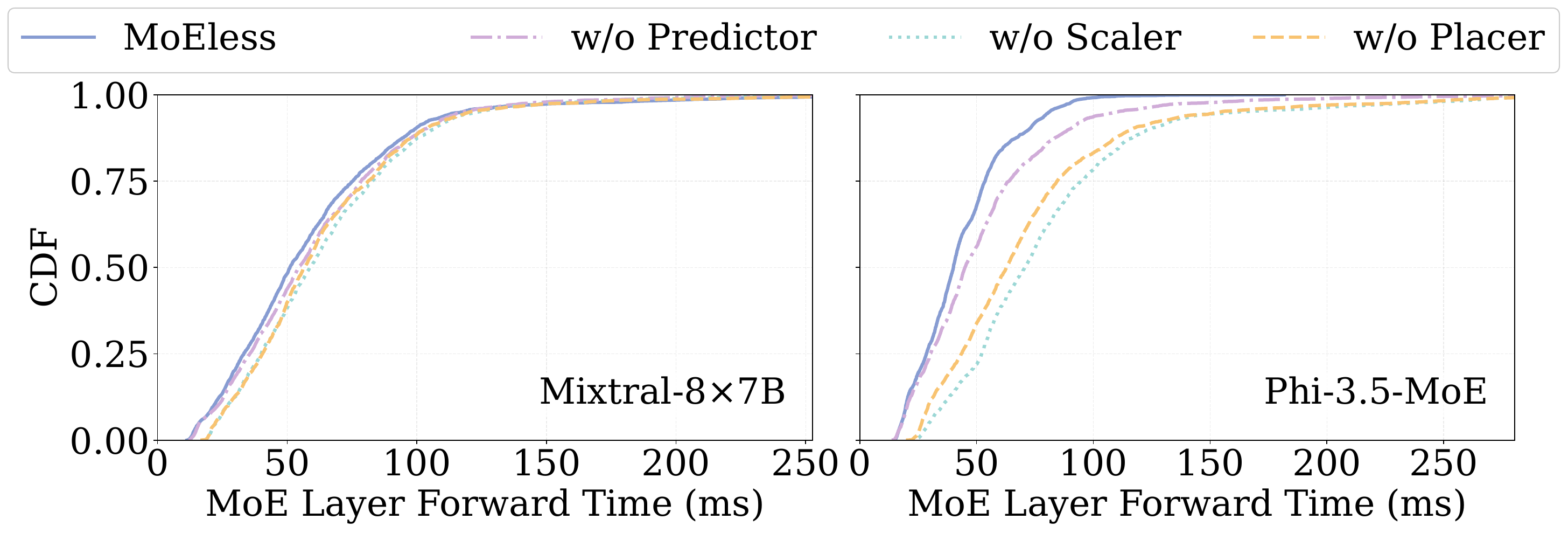}
    \caption{Ablation study of \sys.}
    \label{fig:eval-ablation}
\end{figure}

Figure~\ref{fig:eval-ablation} presents the \CDF of \MoE layer forward latency for \sys and each ablated variant.
Across both models, removing any individual component consistently shifts the latency distribution to the right, confirming that all three modules contribute to the overall performance.
Among them, disabling the Expert Load Predictor incurs the largest performance degradation, as inaccurate load estimation leads to suboptimal expert provisioning and placement decisions.
Removing the Expert Scaler also increases latency by preventing \sys from elastically adjusting the number of serverless experts to match workload demand, resulting in higher queuing delay under load.
Finally, disabling the Expert Placer degrades performance by reducing expert locality and load balancing across GPUs, leading to increased communication overhead and execution imbalance.
Overall, these results demonstrate that the three components are complementary, and together enable the low-latency serving achieved by \sys.

\subsection{System Overheads}
\label{subsec:eval-overheads}

We present a fine-grained breakdown of \sys's overheads, including predictor fine-tuning overhead, predictor memory footprint, and expert operation overhead.

\textbf{Predictor fine-tuning overheads.}
The fine-tuning process is computationally lightweight across three models, as the largest predictor contains only 80K parameters.
Across all three \MoE models, the complete set of predictors can be fine-tuned within five minutes on a single GPU, incurring negligible overhead and enabling rapid deployment.

\textbf{Predictor memory footprints.}
Table~\ref{tab:predictor_footprint} reports the total GPU memory consumption of all predictors across the three models for different methods.
Our predictors occupy approximately 1.50\%, 3.23\%, and 3.19\% of ProMoE's predictor memory footprint for \mixtral, \phimoe, and \llama, respectively.

\textbf{Expert operation overheads.}
The prediction delay is below 0.2~ms per layer, and nearly all expert scaling and placement operations are warm-started without runtime delays.
Moreover, all expert-related operations are executed asynchronously, ensuring minimal impact on inference latency.

\section{Related Work}

\noindent \textbf{Mitigating expert load imbalance in \MoE systems.}
Expert load imbalance is a major performance bottleneck in distributed \MoE systems, as a small subset of popular experts often become stragglers while the remaining experts are underutilized.
Existing approaches address this problem through expert replication, routing optimization, and dynamic expert placement.
FasterMoE~\cite{he2022fastermoe} introduces a dynamic shadowing mechanism to mitigate skewed expert selection during training.
NetMoE~\cite{liu2025netmoe} improves expert routing efficiency by considering the locality of training samples and dynamically rearranging expert placements to reduce all-to-all communication costs.
Lina~\cite{li2023accelerating} alleviates communication bottlenecks during inference through dynamic resource scheduling.
DeepSeek proposes \EPLB~\cite{liu2024deepseek}, which periodically swaps low-usage experts with replicas of popular ones to balance expert loads.
Although effective at reducing expert stragglers, these methods rely on persistent GPU-resident expert replicas, causing GPU memory consumption to increase with replication aggressiveness.
Unlike these serverful approaches, \sys mitigates expert load imbalance through elastic serverless experts.

\noindent \textbf{Optimizing expert parallelism in \MoE training.}
Another line of work improves \MoE efficiency by optimizing expert parallelism during distributed training.
SmartMoE~\cite{zhai2023smartmoe} adaptively combines multiple parallelism strategies to accelerate \MoE training.
ScMoE~\cite{cai2025shortcut} overlaps different forms of parallelism to effectively reduce all-to-all communication latency.
Comet~\cite{zhang2025comet} achieves fine-grained communication--computation overlap by leveraging data dependency analysis and task rescheduling.
MoE Parallel Folding~\cite{liu2025moe} decouples attention and \MoE layers in Transformer blocks, allowing each to independently select the most efficient parallelism strategy.
These systems improve distributed training throughput by reducing communication overhead and improving parallel execution efficiency, but they assume relatively stable expert placement and long-running training workloads.
Orthogonal to these efforts, \sys targets online \MoE serving, where dynamically changing request distributions require elastic expert provisioning rather than optimized training schedules.

\noindent \textbf{Expert offloading for resource-limited \MoE serving.}
Expert offloading reduces GPU memory pressure by dynamically moving experts between GPU, CPU, and storage.
MoE-Infinity~\cite{xue2024moe} traces expert selection patterns to offload inactive experts and prefetch important ones based on predictions.
FineMoE~\cite{yu2025taming} predicts, prefetches, and caches experts to reduce resource pressure.
ProMoE~\cite{song2024promoe} proactively predicts and prefetches experts using neural-network predictors.
DeepSpeed-Inference~\cite{rajbhandari2022deepspeed} performs parameter offloading at the layer level without considering expert awareness.
Mixtral-offloading~\cite{eliseev2023fast} leverages gate network inputs from preceding layers to speculate expert selections.
These approaches primarily optimize memory usage for resource-constrained serving by reducing the number of resident experts.
In contrast, \sys focuses on distributed \MoE serving, where expert load imbalance and communication overhead dominate performance, and dynamically provisions serverless experts to improve both latency and resource efficiency.

\noindent \textbf{\MoE serving in serverless computing.}
Recent works have explored serving \MoE models on serverless platforms to improve resource utilization and reduce inference cost.
Liu \etal~\cite{liu2025optimizing} optimize the deployment cost of \MoE models on serverless platforms using Bayesian optimization.
Remoe~\cite{liu2025remoe} offloads expert modules to CPUs to reduce memory overhead and inference cost on heterogeneous hardware.
sMoE~\cite{yue2026smoe} uses \DRL to configure serverless functions for \MoE serving, which incurs heavy \DRL training overheads.
FaaSMoE~\cite{wang2026faasmoe} leverages serverless functions to share experts in multi-tenant \MoE serving.
These systems leverage serverless infrastructures primarily as a cost optimization mechanism.
In contrast, \sys leverages elastic serverless experts as the core abstraction for expert replication and placement, simultaneously reducing inference latency, expert memory footprint, and serving cost.

\begin{table}[t]
\centering
\caption{Predictor memory footprints across different models and methods.}
\label{tab:predictor_footprint}
\scalebox{1.0}{
    \begin{tabular}{lccc}
    \toprule 
    \textbf{Model} & \textbf{Mixtral-offloading} & \textbf{ProMoE} & \textbf{Ours} \\
    \midrule
    Mixtral-8×7B    & 1.92 MB & 128.32 MB & 1.92 MB \\    
    Phi-3.5-MoE     & 4.16 MB & 128.64 MB & 4.16 MB \\
    Llama-4-Scout   & 3.84 MB &120.48 MB & 3.84 MB \\
    \bottomrule
    \end{tabular}
}
\end{table}

\section{Conclusion}

This paper proposes \sys, an efficient serverless \MoE serving framework that explicitly targets expert load imbalance while accelerating inference through fine-grained serverless expert execution.
\sys employs lightweight, low-overhead predictors to accurately capture incoming expert load distributions and proactively identify expert stragglers.
Guided by these predictions, we design optimized expert scaling and placement strategies that dynamically adjust expert instances across GPUs.
We prototype \sys on top of \megatron and deploy it on an eight-GPU testbed for evaluation.
Experiments with open-source \MoE models and real-world workloads demonstrate that \sys reduces inference latency by up to 43\% and inference cost by up to 84\% compared to \SOTA solutions.

\bibliographystyle{ACM-Reference-Format}
\bibliography{references}

\end{document}